\documentclass[11pt]{JHEP3} % 10pt is ignored!

\JHEPspecialurl{http://jhep.sissa.it/JOURNAL/JHEP3.tar.gz}
\usepackage{epsfig,multicol}
\usepackage{graphicx}
\usepackage{epstopdf}

\usepackage{amsmath}
\usepackage{amssymb}
\usepackage{graphicx,graphics}
\usepackage{amsmath,amssymb,latexsym}
\newcommand{\p}{\partial}
\newcommand{\eq}{\begin{equation}}
\newcommand{\eqe}{\end{equation}}
\newcommand{\nn}{\nonumber}

\newcommand{\eqa}{\begin{eqnarray}}
\newcommand{\eqae}{\end{eqnarray}}

\title{Embedding DBI inflation in scalar-tensor theory}

\author{Carsten van de Bruck\\
{School of Mathematics and Statistics, University of Sheffield,
Hounsfield Road, Sheffield S3 7RH, United Kingdom}}

\author{David F.~Mota\\
{Institute of Theoretical Astrophysics, University of Oslo,
0315 Oslo, Norway}}

\author{Joel M. Weller\\
{School of Mathematics and Statistics, University of Sheffield,
Hounsfield Road, Sheffield S3 7RH, United Kingdom}}

\preprint{}

\abstract{
The Dirac-Born-Infeld (DBI) action has been widely studied as an interesting example of a
model of k-inflation in which the sound speed of the cosmological perturbations differs from unity.
In this article we consider a scalar-tensor theory in which the 
matter component is a field with a DBI action.
Transforming to the Einstein frame, we explore the effect of the resulting coupling on the
background dynamics of the fields and the first-order perturbations. 
We find that
the coupling forces the scalar field into the minimum of its 
effective potential. While the additional scalar field contributes significantly to the energy density during inflation, the dynamics are determined by the DBI field, 
which has the interesting effect of 
increasing the number of efolds of inflation and decreasing the boost factor of the DBI field.
Focusing on this case, we show, with the benefit of numerical examples,
that the power spectrum of the primordial perturbations is determined by the behaviour of the perturbations of the modified DBI field.
}
              
\begin{document}

\section{Introduction}

The popularity of string/M theory and its interpretation as a truly fundamental model
of physical phenomena has motivated the search for a mechanism to explain the 
process of inflation. In canonical models of inflation,
 a slowly rolling scalar field, with its energy density dominated by its potential, is introduced 
as a phenomenological tool to account for the dramatic expansion of the universe to solve the flatness and horizon problems of the Big Bang theory and also as 
a source of the primordial fluctuations that give rise to structure formation and the observed
anisotropies in the CMB radiation. 
While there is no shortage of scalar fields in stringy models, problems arise when one looks for a suitable candidate to play the role of inflaton in the canonical models  \cite{McAllister:2007bg, Baumann:2009ni}. The abundance of light scalars (moduli) not only complicates the dynamics of inflation, but also means that models must be tuned to prevent unwanted light moduli affecting the post-inflationary universe. As well as this, although heavy fields are effectively frozen during inflation they may contribute to the potential that determines the evolution of the dynamical fields,
 further complicating the `eta problem' in inflationary model building, in which one encounters large corrections to the flat potential required in slow-roll inflation \cite{Copeland:1994vg,Stewart:1994ts,eta1, Easson:2009kk,eta3}.

An intriguing scenario is that of Dirac-Born-Infeld (DBI) inflation \cite{Silverstein:2003hf,Alishahiha:2004eh} in which the role of inflaton is played by the radial coordinate of a D3 brane moving in a warped region (throat) of a compactified space. Crucially, a speed limit is imposed upon of the motion of the brane (which appears as a point-like object in the compact dimensions), affected by both its speed and the warp factor of the throat. The speed limit introduces a parameter $\gamma$, which plays an analogous role to the Lorentz factor in special relativity, that grows without bound as the speed limit is approached.  This has the effect that in the 'relativistic' regime (as we shall call it), the importance of the friction and potential terms in the scalar field equation of motion is lessened, so that a DBI field can drive inflation with a relatively steep potential. 
 
From a cosmological point of view, DBI inflation is important as it is a concrete example of k-inflation i.e. inflation driven by a scalar field with a non-canonical kinetic term \cite{ArmendarizPicon:1999rj, Garriga:1999vw,k2,k3,k4,k5,k6,k7,k8}. A distinguishing feature of these models is that the field perturbations travel at a speed less than that of light, with the result that the Fourier modes freeze in at a sound horizon that is different from the curvature radius and which changes dynamically with the background evolution \cite{Garriga:1999vw}. This is an exciting possibility, as the amplitude of the primordial perturbations would not be (almost) independent of wavenumber as in the standard inflationary models and the effects of the evolution of the universe during the inflationary regime could be encoded in mode correlations. In DBI inflation, the sound speed is given by $\gamma^{-1}$, and decreases to small values as the speed limit is saturated. Observational limits on the level of non-Gaussianities in the CMB radiation can thus be combined with traditional constraints arising from the power spectrum of primordial perturbations to distinguish between DBI and canonical models of inflation (see \cite{Bartolo:2004if,Liguori:2010hx} for recent reviews).

It is to be expected that the effective action governing the inflationary phenomenon will contain other features characteristic of fundamental theories such as non-minimal coupling and multiple fields. In particular,
the effect of the presence of
many scalar fields on the inflationary dynamics has been the subject of intense study over the past decade, with many works focusing on the behaviour of entropic or isocurvature perturbations in these models. These can source the adiabatic fluctuations in the system so that the curvature perturbation, which in single field models is conserved outside the horizon, acquires a dependence on the inflationary dynamics. 
Interestingly, the presence of a coupling of the type $A(\phi)=\exp(\beta\phi)$ 
between two scalar fields (arising naturally in scalar-tensor theories considered in the Einstein frame)
has been found to further influence the transfer between entropic modes and adiabatic modes \cite{DiMarco:2002eb,Lalak:2007vi}; however,
the analogous case for multifield k-inflation models has been neglected in the literature.
Multiple fields arise naturally in DBI
scenarios, where the extra degrees of freedom can be identified with the non-radial coordinates of the D3 brane in the compactified space, and several studies (both analytical and numerical) have been devoted to understanding the  
evolution of adiabatic and isocurvature perturbations and the predicted level of non-Gaussianity in these models \cite{dbi10,dbi9,dbi1,dbi3,dbi2,dbi7,dbi4,dbi8}.
In contrast, the question of the behaviour of the perturbations in models involving many fields, each with a different intrinsic sound speed, has received relatively little attention. One such model was investigated in \cite{Cai:2009hw},
in which inflation is driven by two standard DBI fields, each corresponding to a brane with its own sound speed dependent on its evolution. The presence of two sound speeds means that the perturbations freeze in at the sound horizon corresponding to the larger value of $c_s$ while the non-Gaussianities are at a level corresponding to the square of the larger value of $\gamma$. 

In order to address these issues, we propose a model in which a scalar field with a DBI action evolves in the presence of a canonical scalar field. We embed the DBI inflationary scenario into a scalar-tensor theory, in which the additional field could describe the degree of freedom associated with additional moduli fields in the higher-dimensional theory. We will not construct such a model from first principles, but rather treat it as a phenomenological model. Therefore, we consider the generalised case in which the fields are coupled, motivated by the possibility of the additional scalar field having a non-minimal coupling to gravity in the effective action. 
Notwithstanding the importance of non-minimal coupling from a fundamental theory standpoint, couplings of this type have often been neglected in the DBI inflation literature, partly due to the success of the standard DBI scenario as an consistent phenomenological model in their absence.
However, we should point out that there have been some interesting studies such as \cite{Easson:2009kk}, where non-minimally coupled DBI models were considered in order to circumvent the eta problem in inflationary model building. Some issues pertaining to the realisation of non-minimal coupling in DBI models of inflation from the string theory perspective were raised in \cite{Silverstein:2003hf} and also \cite{Kachru:2003sx}. 
The authors of \cite{Easson:2009wc} performed a detailed analysis of a single field DBI scenario with non-minimal coupling of the Brans-Dicke type. Their analytical work shows that both in the small $\gamma$  and large $\gamma$ limits the effect of the coupling is negligible but in the intermediate regime, the degeneracy of observables can be broken: for a given tensor-to-scalar ratio and medium $f_{NL}$, non-minimally coupled DBI models have a redder scalar spectral index.  
The effect of coupling in multifield DBI models was also considered in \cite{Brax:2009hd,Brax:2010tq}, focusing on the effect of particle production due the interaction between the 'inflaton' brane and trapped branes in the warped throat. Although,
like our model, this scenario involves a DBI action non-minimally coupled to a second field, the motivation and dynamics are very different: for example, in our model the scalar field as well as the DBI field can contribute significantly to the energy density and the background dynamics depend strongly on the evolution of its vacuum expectation value (VEV). Recently, coupled DBI inflation has also been considered in the context of  a warm inflationary model \cite{Cai:2010wt}.

This paper is organised as follows. In Sec. \ref{Sec:BGround} we introduce the model in the 
Jordan and Einstein frames, and derive the equations of motion for the inflationary Universe.
Sec. \ref{Sec:Dynamics} is devoted to an analysis of the background dynamics, both in the 
non-minimally coupled and coupled cases and in Sec.  \ref{Sec:Perturbations} the 
equations determining the evolution of the first-order perturbations are derived 
and solved numerically.
We conclude in Sec. \ref{Sec:Conclusions} with a short discussion.

\section{The Model}
\label{Sec:BGround}

Non-minimally coupled scalar fields arise in a number of scenarios in high-energy physics, such as 
low-energy effective actions from higher dimensional theories, quantum field theory in curved space and $f(R)$ models of gravity. We consider a scalar-tensor theory with canonical field $\varphi$ coupled to the Ricci scalar combined with a second field $\chi$ with a DBI action. In the Jordan frame, this
is described by the action
\eq
S=\int d^4 x \sqrt{-g} \left[ F(\varphi)\frac{R}{2}-\frac{1}{2}(\nabla\varphi)^2-U(\varphi)+
f^{-1}(\chi)\left[  1-\gamma^{-1} \right]-V(\chi) \right], 
\label{eq:JordanAction}
\eqe
where the function $F(\varphi$) determines the coupling of the canonical scalar field to gravity, $f(\chi)$
is the warp factor of the DBI field, $U(\varphi)$ and $V(\chi)$ are potentials for the canonical and DBI fields 
respectively and $\gamma$ is the boost factor of the DBI field, given by
\eq
\gamma=\frac{1}{\sqrt{ 1+f(\chi)(\nabla\chi)^2 } }.
\eqe
Varying (\ref{eq:JordanAction}) with respect to the metric gives the modified Einstein equations,
\eq
F G_{\mu\nu}= T^{\textrm{DBI}}_{\mu\nu} +[\p_\mu\varphi\p_\nu\varphi -\tfrac{1}{2}g_{\mu\nu}(\nabla\varphi)^2 ]
-g_{\mu\nu}U+\nabla_\mu\nabla_\nu F-g_{\mu\nu}\Box F,
\eqe
where $\Box\equiv g^{\mu\nu}\nabla_\mu\nabla_\nu$ is the d'Alembertian operator and $T^{\textrm{DBI}}_{\mu\nu}$ the energy momentum tensor of the DBI field. The equations of motion for the fields are 
\eqa
2\varpi\Box\varphi = F'T-\varpi' (\nabla\varphi)^2 - 4UF' + 2U'F~, \\
\nabla_\mu\left[\gamma g^{\mu\nu}\nabla_\nu\chi\right] = \frac{f'}{f^2}(1-\tfrac{1}{2}\gamma-\tfrac{1}{2}\gamma^{-1})+V'~,
\eqae
where $\varpi\equiv F+\tfrac{3}{2}F'^2$. In the Jordan Frame, the equations governing the dynamics are evidently quite complicated. 
In order to simplify the system and gain a different perspective we can perform a conformal transformation
to the Einstein frame in which the Hilbert term in the action takes the standard form.
We define 
\eq
\tilde{g}_{\mu\nu}=F(\varphi)g_{\mu\nu},
\eqe
where the tilde denotes quantities evaluated in the new frame in which the $\varphi$ field is minimally coupled to gravity. Using the definitions,
\eq
\frac{d\tilde{\varphi}}{d\varphi}=\sqrt{ \frac{3}{2}\left(\frac{F_\varphi}{F}  \right)^2+\frac{1}{F} },
\eqe
\eq
\tilde{U}(\tilde{\varphi})=U(\varphi)F^{-2},
\eqe
\eq
A(\tilde{\varphi})=F^{-\frac{1}{2}},
\eqe
\eq
\tilde{\gamma}=\frac{1}{\sqrt{ 1+A^{-2}f(\chi)\tilde{g}^{\mu\nu}\chi_{,\mu}\chi_{,\mu} }},
\eqe
the action in the Einstein frame can be written as
\eq
S=\int d^4 x \sqrt{-\tilde{g}} \left[ \frac{\tilde{R}}{2}-\frac{1}{2}\tilde{g}^{\mu\nu}\tilde{\varphi}_{,\mu}\tilde{\varphi}_{,\mu}
-\tilde{U}(\tilde{\varphi}) \right]+S_{\mathrm{DBI}}\left[ A^2 \tilde{g}_{\mu\nu}  \right]
\label{eq:Einsteinaction}
\eqe
where,
\eq
S_{\mathrm{DBI}}\left[ A^2 \tilde{g}_{\mu\nu}  \right]=
\int d^4 x \sqrt{-\tilde{g}} A^4\left\{   f^{-1}(\chi)\left(1-\tilde{\gamma}^{-1}\right)-V(\chi) \right\}.
\eqe
The scalar field can thus be redefined to retain its canonical kinetic term in this frame; the DBI field
is not redefined but its behaviour will be affected by the coupling terms $A(\tilde\varphi)$ present both
as an overall multiplier in the action but also under the square root in the definition of the 
boost factor $\tilde\gamma$. Throughout the rest of this paper we shall work in the Einstein frame and so will neglect the tildes for convenience of notation.

We study the flat FRW metric in this frame, describing an expanding universe, 
\eq
ds^2 = -dt^2 + a^2(t)\delta_{ij}dx^i dx^j~,
\eqe
 with scale factor $a(t)$. The background equation of motion for the homogeneous field $\chi$ is,
 \eq
 \ddot\chi+3 H\gamma^{-2}\dot\chi + \tfrac{1}{2}A^2\frac{f_\chi}{f^2}(1-3\gamma^{-2}+2\gamma^{-3}) + A^2\gamma^{-3}V_{\chi}=
-\beta\dot\chi\dot{\varphi}(3\gamma^{-2}-1),
\label{eq:Echi}
 \eqe
where dots indicate derivatives with respect to the time coordinate in the Einstein frame, $f_\chi\equiv df/d\chi$, $H = \dot{a}/a$ and $\beta=d\ln A/d\varphi$. Throughout this paper we will assume that $\beta$ is constant, so the coupling $A(\varphi)$ is given by
\eq \label{eq:Adef}
A(\varphi) = \exp(\beta\varphi).
\eqe
Eqn. (\ref{eq:Echi}) reduces to the standard DBI equation of motion when $\beta=0$, although it is important to note that in the coupled case, the boost factor $\gamma$ acquires a factor of $A$ under the square root that affects the dynamics of the field,
\eq \label{eq:gam}
\gamma = \frac{1}{\sqrt{1-A^{-2}f\dot\chi^2}}.
\eqe
The $\varphi$ equation takes the simple form,
\eq \label{eq:varphi}
\ddot{\varphi}+3H\dot{\varphi}+U_{\varphi}=\beta T_{\textrm{DBI}},
\eqe
in terms of the trace of the stress-energy tensor of the DBI field
 $T_{\textrm{DBI}}=-\rho_{\textrm{DBI}}+3p_{\textrm{DBI}}$ . 
 The energy density and pressure for $\chi$ are
\eq
\rho_{\textrm{DBI}}=A^4\left[f^{-1}(\gamma-1)+V\right] \hspace{1.5cm} 
p_{\textrm{DBI}}=A^4\left[f^{-1}(1-{\gamma}^{-1})-V\right].
\eqe
so $T_{\textrm{DBI}}$ can be written in terms of $\gamma$, $A$ and $V$ as
\eq
 T_{\textrm{DBI}}= A^4 \left[ f^{-1}( 4-3\gamma^{-1}-\gamma )-4V \right].
\eqe 
The Friedmann equations are
\eqa
3H^2 &=& \tfrac{1}{2}\dot{\varphi}^2+U+ \rho_{\textrm{DBI}}, \\
-2\dot{H} &=& \dot{\varphi}^2+\gamma A^2\dot\chi^2. \label{eq:Fried2}
\eqae
From these we can define the slow-roll parameter, $\epsilon$,
\eq \label{eq:epsilon}
\epsilon\equiv -\frac{\dot{H} }{H^2}=\frac{3}{2}\frac{ \dot{\varphi}^2+\gamma A^2\dot\chi^2}{ \rho_{\textrm{DBI}}+\tfrac{1}{2}\dot{\varphi}^2+U},
\eqe
in the standard way, so that $\ddot{a}/a = H^2 (1-\epsilon)$.

\section{Background Dynamics}
\label{Sec:Dynamics}
\subsection{DBI inflation}

\begin{figure}[t!] 
\centering
\includegraphics[width=1\textwidth]{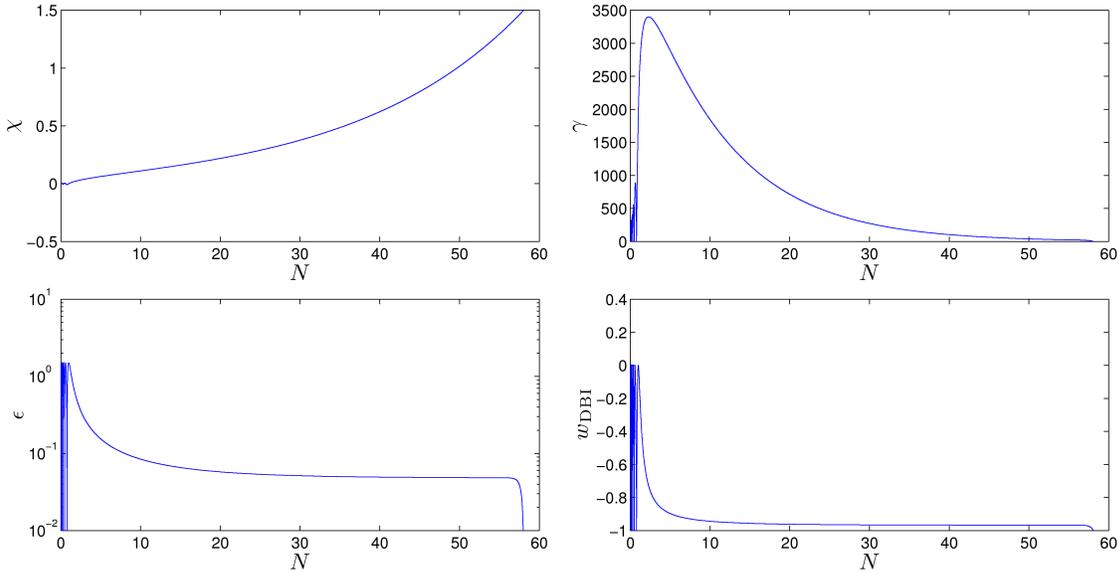}
\caption{DBI inflation, plotted against $N=\ln a$. The upper left panel shows the $\log$ of the field value of $\chi$; the upper right panel shows the increase in the boost factor $\gamma$. The lower left panel shows the logarithm of the slow-roll parameter $\epsilon$, and the bottom right shows the equation of state parameter $w_{DBI} = p/\rho$, which remains close to $-1$ until $\chi$ starts to oscillate about its minimum. A reasonable number of efolds of inflation are generated in this case, but the boost factor gets very large. Parameter Values: $\lambda=10^{12}$, $m=5\times 10^{-5}$, $\mu=0.1$. }
\label{fig:DBIonly_N}
\end{figure}

Before discussing the two-field system, it will be useful to remind the reader of some of the features of single-field DBI inflation \cite{Silverstein:2003hf}.
Setting $\beta = 0$ in (\ref{eq:Echi}) the equation of motion for the minimally coupled DBI field can be written as
\eq
\ddot\chi+3H\gamma^{-2}\dot\chi+\tfrac{1}{2}\frac{f_\chi}{f^2}(1-3\gamma^{-2}+2\gamma^{-3})+\gamma^{-3}V_\chi =0.
\eqe
As the $\chi$ field rolls down its potential, $\gamma$ gradually increases from 1, driven by the potential.   
As $|\dot\chi|$ increases, so does the boost factor, with the result that terms in $1/\gamma$ are suppressed. The potential and friction terms then drop out and the equation reduces to,
\eq \label{eq;DbiSimp}
\ddot\chi+\tfrac{1}{2}\frac{f_\chi}{f^2} \approx 0 \ \ \Rightarrow\ \  \ddot\chi-\frac{2}{\lambda}\left( \chi^3 + \mu^2\chi \right) \approx 0,
\eqe
where in the second equation we have used the standard 'mass gap' form for the warp factor, given by
\eq
f(\chi) = \frac{\lambda}{(\chi^2+\mu^2)^2}.
\eqe
This is an approximation to the Klebanov-Strassler throat (see \cite{Kecskemeti:2006cg} and references therein) used to describe the warped conifold in which the DBI field moves that reproduces phenomenologically the form of $f(\chi)$ in both the IR and UV limits. 
In the limit of an AdS throat ($\chi^2\gg \mu^2$) the solution is $\chi\sim\sqrt{\lambda}/t$; towards the tip of the throat where $\chi^2\ll \mu^2$, $\chi$ decreases faster as $\chi\sim\exp(-\mu\sqrt{2/\lambda}t)$. 
 
 The slow-roll parameter $\epsilon$ is given by,
\eq
\epsilon \equiv -\frac{\dot{  H}}{H^2} = \frac{3}{2}\frac{f^{-1}(\gamma-\gamma^{-1})}{f^{-1}(\gamma-1)+V}
\approx   \frac{3}{2}\frac{\gamma}{\gamma+fV}~.
\eqe
Note that although $\gamma$ gets large, the large value of $f$ required to satisfy observational constraints \cite{Alishahiha:2004eh} means that the potential term is the dominant contribution to the energy density and the pressure, implying that $\epsilon$ is less than 1 when $\chi$ is small and the equation of state $w_{\rm DBI} = p_{\rm DBI}/\rho_{\rm DBI} $
\eq
w_{\rm DBI} = \frac{f^{-1}(1-{\gamma}^{-1})-V}{f^{-1}(\gamma-1)+V}
\approx  -\frac{fV}{\gamma+fV}
\eqe
 becomes very close to $-1$.  This is plotted in Fig. \ref{fig:DBIonly_N}. 
 
 Single field DBI inflation is normally treated analytically using the Hamilton-Jacobi formalism, in which the field value $\chi$ assumes the role of the time-coordinate. Differentiating $H=H(\chi)$ with respect to cosmic time and using the second Friedmann equation, one finds,

\begin{equation} \label{eq:HJ1}
\dot\chi =   -2\frac{H'(\chi)}{\gamma(\chi)}
\end{equation}
Therefore, the definition of $\gamma$ can be rewritten as,
\begin{equation} \label{eq:HJ2}
\gamma(\chi) = \sqrt{1+4f(\chi)H'(\chi)^2},
\end{equation}
and the potential,
\begin{equation} \label{eq:HJ3}
V(\chi) = 3H(\chi)^2-f(\chi)^{-1}(\gamma(\chi)-1).
\end{equation}
The slow-roll parameter
\footnote{
One can generally construct a 'tower' of flow (slow-roll) parameters  (the first two being $\epsilon$ and $\eta$) defined in terms of derivatives of $H(\chi)$ with respect to $\chi$ \cite{Kinney:2002qn}. In addition, it has been shown in \cite{Tzirakis:2008qy} that one can additionally construct a second 'tower' in terms of derivatives of $\gamma(\chi)$ with respect to $\chi$. The first flow parameter in this 'tower', $s$, enters the expression for the spectral index of curvature perturbations as a measure of the variation of the speed of sound i.e. $n_s-1 = -4\epsilon+2\eta-2s$.
}
can then be written
\begin{equation} \label{eq:DBIepsilon}
\epsilon = \frac{2}{\gamma}\left(\frac{H'}{H}\right)^2 \approx \sqrt{\frac{3}{\lambda}}\frac{1}{m},
\end{equation}
where the second equality holds for the AdS throat with potential $m^2\chi^2$ when $\epsilon\ll1$. An unusual feature of DBI inflation is that an increased mass term decreases $\epsilon$, unlike in the standard slow-roll case. The advantage of the Hamilton-Jacobi approach is that, given a potential $V(\chi)$,  eqns. (\ref{eq:HJ3}) (using the definition of $\gamma(\chi)$ in  (\ref{eq:HJ2})) and (\ref{eq:HJ1}) can be solved sequentially. One can also choose an anzatz such as $H(\chi)\propto\chi$ in order to calculate $V$. 

In much of the DBI literature $m^2\chi^2$ potentials similar to those in large field inflation are used, following the analysis in \cite{Silverstein:2003hf}. This showed not only could a mass term be generated in the DBI scenario without the setup being ruined by back-reaction of the probe brane, but it could drive the power-law inflation discussed above.  
Although it has since been found \cite{Spalinski:2007dv,Chimento:2007es, Bessada:2009pe} that the
class of potentials that lead to successful DBI inflation is much wider, in this work we focus on modifications to the standard DBI scenario and will restrict ourselves to the potential
\eq
V = \tfrac{1}{2}m^2\chi^2.
\eqe
In the cutoff throat, the $\gamma$ increases to a maximum then decreases slightly as the friction term starts to become more important. Generally, $\gamma$ remains large even after $\chi$ has passed through the origin, whereupon the other terms in the equation start to combine to stop the field, giving $\gamma=1$. After this point, the field exhibits oscillations about its minimum. 
The presence of one or more $\overline{\mathrm{D}3}$ branes in the tip region of many throat constructions, means that DBI inflation is often taken to end with the probe brane annihilating at the point $\chi_{\mathrm{end}}$, rather than with a violation of the condition for acceleration $\epsilon<1$. The mechanism by which the energy released in this scenario is transferred to the standard model brane is discussed in \cite{Barnaby:2004gg}. If $\chi$ can reach the bottom of its potential, it can oscillate, which allows for the possibility of energy transfer in a period of (p)reheating \cite{Lachapelle:2008sy, Bouatta:2010bp}, however, this is beyond the scope of the present paper.
 
As discussed in the introduction, DBI inflation is a particular case of k-inflation in which the squared sound speed of scalar perturbations is equal to $c_s^2=\gamma^{-2}$.
The non-Gaussianity of the primordial perturbations is determined by the bispectrum (the three-point
correlation function of the scalar curvature perturbation $\zeta$), which in general is a 
non-trivial function of the momentum configuration.
The bispectrum can be decomposed into several parts (shapes), each with a magnitude parameterised
by an estimator (or non-linearity parameter) $f_{NL}$ defined by an ansatz of the form  \cite{Chen:2006nt}
\begin{equation}
\zeta = \zeta_L-\frac{3}{5}f_{NL}\zeta_L^2,
\end{equation}
where $\zeta_L$ is the linear part of the scalar curvature perturbation. 
In DBI inflation, the leading contribution is estimated by\footnote{
Eqn. (\ref{eq:fNL_DBI}) is calculated on an equilateral triangle using momenta of equal amplitude.
This is relevant because the scalar perturbations are frozen on large scales and oscillate (averaging out)
on small scales. Thus, the dominant contribution comes from modes with similar wavelengths that exit the
(sound) horizon at the same time. 
}
\begin{equation} \label{eq:fNL_DBI}
f_{NL}=\frac{35}{108}\left( \frac{1}{c_s^2}-1 \right).
\end{equation}
Therefore,
\begin{equation}\label{eq:fNL}
f_{NL}\approx 0.32\gamma^2.
\end{equation}
Conservative estimates of the observational value\footnote{
We note in passing that the measured value of $f_{NL}$ can from receive relevant corrections from the
post-inflationary evolution of the perturbations
and differ from the expressions given above. \cite{Bartolo:2004if}. The curvature perturbation, however, is 
conserved on large scales so one can connect the observable signal to the primordial perturbations.  
}
of the non-linearity parameter have $|f_{NL}|\lesssim \mathcal{O}(100)$, leading to an upper bound on the value of the boost factor $\gamma\lesssim 20$, assumed to apply during the first 10 of the last 60 efolds of inflation which directly affect the CMB radiation. Since the boost factor increases with the number of efolds of inflation $N_{\rm max}$, many DBI models with $N_{\rm max}$ sufficiently large predict a high level of non-Gaussianity exceeding current observational bounds (cf. \cite{Bean:2007hc,Peiris:2007gz} and also \cite{Alishahiha:2004eh,Baumann:2006cd,Kecskemeti:2006cg}).

\subsection{Minimal Coupling $\beta=0$}

\begin{figure}
\centering
\includegraphics[width=1\textwidth]{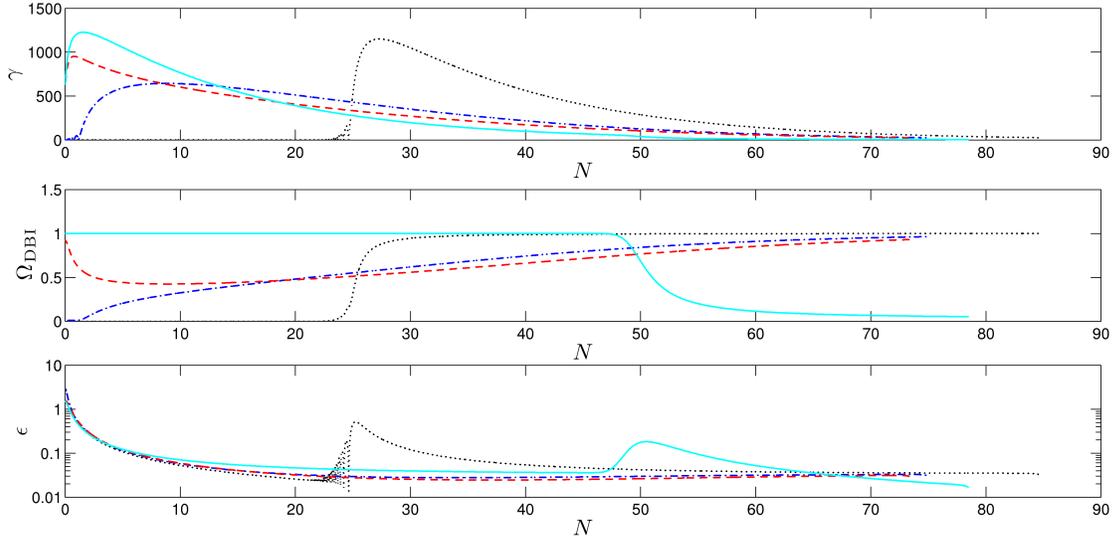}
\caption{Minimal coupling with a quadratic potential $U(\varphi)=U_0\varphi^2$ with $U_0 = 10^{-14}$ (black, dotted), $10^{-12}$ (blue, dot-dashed), $2\times10^{-12}$ (red, dashed) and $5\times10^{-10}$ (cyan, solid). The upper panel shows the boost factor $\gamma$, the middle panel show the  DBI density parameter $\Omega_{\rm DBI}=\rho_{\rm DBI}/3H^2$ and the lower panel the slow-roll parameter $\epsilon$. Plots are against efold number $N=\ln({\rm max}(a)/a)$ so $N=0$ corresponds to the end of inflation.
Other parameter values:  $\lambda=2\times10^{12}$, $m=5\times 10^{-5}$ and $\mu=0.2$.
 }
\label{fig:MinCoup1}
\end{figure}

\begin{figure}
\centering
\includegraphics[width=1\textwidth]{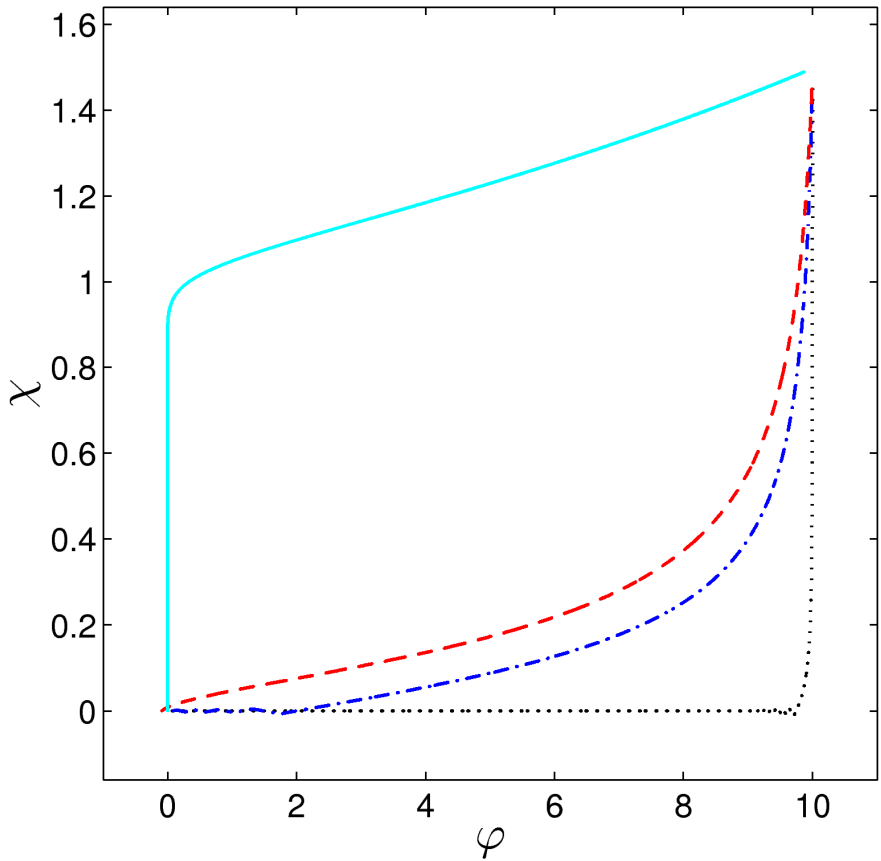}
\caption{The evolution of the canonical and DBI fields in the $(\varphi,\chi)$ plane for the plots shown in fig. \protect\ref{fig:MinCoup1}.}

\label{fig:MinCoup2}
\end{figure}

Let us now include the $\varphi$ field.  A priori, there appear to be five possible ways in which the system can behave, viz. $\chi$ dominated inflation, $\varphi$ dominated inflation, a period of double inflation with a changeover from $\chi$ to $\varphi$ (or vice-versa) or a setup in which $\chi$ and $\varphi$ contribute roughly equally to the energy density.

In the minimally coupled case, the fields $\chi$ and $\varphi$ only interact gravitationally and since  they evolve on different time scales in this case, it is difficult to get a situation in which both contribute equally to the energy density for a considerable length of time. 
Even if one fine-tunes the initial conditions so the fields have similar energy densities, there are problems. $\chi$ becomes small quickly unless its potential is large, but if we require  $\Omega_\chi \approx \Omega_\varphi$, 
 (where $\Omega_i = \rho_i/3H^2$ is the density parameter for the i$^{\rm th}$ field) the potential for $\varphi$ is steep and the field quickly rolls to its minimum.  The opposite is true if $V\approx U$ is smaller: since the DBI contribution to the slow-roll parameter 
 depends explicitly on the DBI field mass, it decays quickly, leaving $\varphi$ to slowly roll to its minimum. 

 If the total energy density is dominated by the DBI potential, the evolution of $\chi$ proceeds in the usual fashion: $\gamma$ increases steadily to large values, until $\chi \ll \mu$, whereupon the field exhibits oscillations. If the difference in energy densities between the two fields in not great, the $\varphi$ field rolls in the background. If, however, $3H^2 \approx V(\chi)$, Hubble damping of the $\varphi$ field is disproportionally large, so the field
evolves slowly until the DBI field decays. When this occurs, the $\varphi$ field rolls to its minimum as in canonical slow-roll inflation. An example of this behaviour is the black (dotted) curve in figs. \ref{fig:MinCoup1} and \ref{fig:MinCoup2}.

The opposite occurs when the $\varphi$ field is initially dominant (i.e. , $3H^2 \approx U(\varphi)$): one finds a period of standard slow-roll expansion followed by DBI inflation (cf. solid cyan curve in figs. \ref{fig:MinCoup1} and \ref{fig:MinCoup2}). At first $\Omega_{\varphi}>\Omega_{\chi}$ and $\gamma\sim 1$ so the DBI field is dynamically unimportant, simply frozen or evolving slowly to smaller values. The timescale for $\varphi$ inflation is rapid compared to that of DBI inflation so $\gamma$ does not increase much during this stage if $\chi$ is rolling. When the $\varphi$ field decays, the slow-roll parameter grows, only to decrease again as the boost factor increases. In both cases, if one field is dominant, there are two distinct periods of inflation.

Perturbations in two-field inflationary models can often be decomposed into modes tangential (adiabatic) and orthogonal (entropic)  to the 
background trajectory in field space \cite{Gordon:2000hv, DiMarco:2002eb,GrootNibbelink:2001qt,Lalak:2007vi}.
The trajectories for the two aforementioned examples exhibit a sharp turn in field space (fig. \ref{fig:MinCoup2}) as the dominant field reaches its minimum. 
In canonical scalar field models, behaviour of this type is of particular interest as it can lead to a 
transfer between entropic (isocurvature) and adiabatic modes, so that the
curvature perturbation is not constant on super-Hubble scales. 
If one fine-tunes the mass scales so that both fields contribute roughly equally to the total energy density for a considerable number of efolds, the dynamics of the DBI field can be significantly affected by the additional contribution to the Hubble damping, leading to smaller values of $\gamma$ (relative to the DBI dominant case). 
As both fields are evolving, the interplay between the field perturbations suggests that
isocurvature fluctuations could be important.
However, minimally coupled case is extremely sensitive to small changes in the field masses (as can be seen in fig. \ref{fig:MinCoup2}) and the initial conditions of both fields so will be analysed in more detail in future work. In the rest of this paper we will focus on the non-minimally coupled case, in which the $\varphi$
field moves in an effective potential dependent on the DBI field.

\subsection{Non-minimal Coupling}
\label{sec:NonMinCoup}

If we consider the non-minimally coupled model interactions between the fields can become more complex as the extra mass term $-\beta T_{\rm DBI}$ in (\ref{eq:varphi}) starts to play an important role. 
The combined effect of this and the bare potential means that the effective potential $U_{\rm eff}=U-\tfrac{1}{4}T_{DBI}$ can have a minimum value at which the
field value $\varphi=\varphi_{\rm min}$ satisfies,
\eq \label{eq:MinCond}
\frac{dU}{d\varphi}\bigg|_{\varphi=\varphi_{\rm min}}-\beta e^{4\beta\varphi_{min}} T_{DBI}^{b}=0,
\eqe 
where $T_{DBI}^{b} = f^{-1}(4-\gamma-3\gamma^{-1})-4V \approx -4V$ is the trace of the `bare' DBI stress-energy tensor and the $\varphi$ dependence of $A(\varphi)$ has been written explicitly using (\ref{eq:Adef}).

The existence of a minimum is dependent on the form of the bare potential: the exponential form of the coupling means that it is necessary to have a potential that slopes to the right\footnote{
Alternatively, if $\beta$ were defined to be negative, for a minimum to exist one would need a potential with positive slope $dU/d\varphi < 0$. In this case, one would need negative values of $\varphi$ to ensure that $A>1$, but the phenomenology of the model would be unchanged.
}
i.e $dU/d\varphi < 0$. 

\subsubsection{Offset quadratic potential}

In this paper, we will consider two choices of potential, dependent on two free parameters $U_0$ and $\eta$. First, the offset quadratic potential
\eq
U(\varphi) = U_0 (\varphi-\eta)^2,
\eqe
for which the true minimum is $\varphi =  \eta$. (The advantages of the offset potential over the standard quadratic $\eta=0$ case will become apparent.) In this case  $dU/d\varphi = 2U_0(\varphi-\eta)$; 
writing  $v=-4\beta \varphi_{min}$, the condition for the minimum is
\[
ve^v= - \frac{2\beta^2  T_{\rm DBI}^{b}e^{4\beta\eta}}{U_0}\approx   \frac{8\beta^2 V(\chi) e^{4\beta\eta}}{U_0},
\]
where the approximation is valid as, generally, the pressure and energy density of the DBI field are dominated by its potential.
$\varphi_{min}$ can then be expressed in terms of Lambert's W function $W(x)$ (satisfying $W(x)e^{W(x)}=x$) as
\eq \label{eq:QuadMin}
\varphi_{min}\approx\eta-\frac{1}{4\beta}W\left( \frac{8\beta^2 V e^{4\beta\eta}}{U_0} \right).
\eqe

\begin{figure}[t]
\centering
\includegraphics[width=1\textwidth]{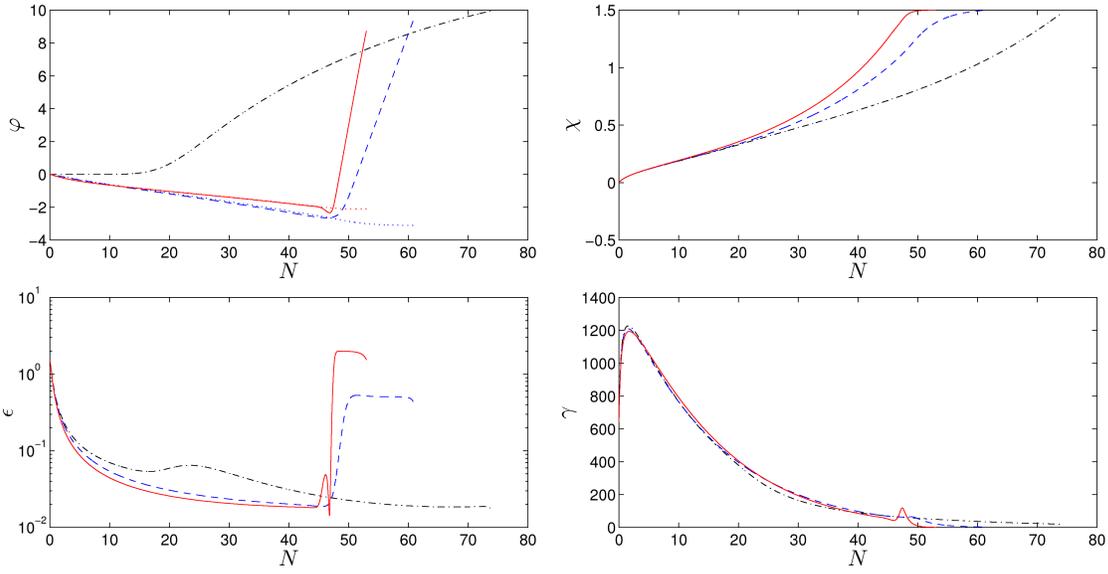}
\caption{
Non-minimal coupling with $A\lesssim1$ for the quadratic potential. Black dot-dash curves have $\beta=0$, blue dashed curves have $\beta=0.25$ and solid red curves have $\beta=0.5$.  
Note that in the latter case the effective potential is steep and there is a brief kination stage during which there is no inflation ($\epsilon>1$) as $\dot{\varphi}$ is large. Parameter values: $\lambda = 2\times 10^{12}$, $m=5\times 10^{-5}$, $U_0 = 2\times 10^{-11}$, $\mu=0.2$, $\eta=0$. 
}
\label{fig:LargeMAlp}
\end{figure}

The effect of the coupling on the dynamics of the system depends strongly on the sign of $\varphi$, as this affects whether the coupling $A=\exp(\beta\varphi)$ is greater or less than one.
Even if $\varphi$ is initially positive, the field will be attracted to its minimum, which can take negative values if $\eta$ is small. As can be seen in fig. \ref{fig:LargeMAlp}, while $\varphi$ is being dragged to smaller values
the extra friction term on the RHS of 
(\ref{eq:Echi}), which is approximately  $\beta\dot\varphi\dot\chi$, acts to further slow the rate of decrease of $\chi$.
When $\varphi$ reaches its minimum, depending on the steepness of the effective potential it will either overshoot (which in extreme cases can lead to values of $A$ substantially less than one) or track the minimum value, which evolves slowly  toward $\eta$ at a rate determined by the decrease in $V(\chi)$.

\begin{figure}[t]
	\centering
	\includegraphics[width=1\textwidth]{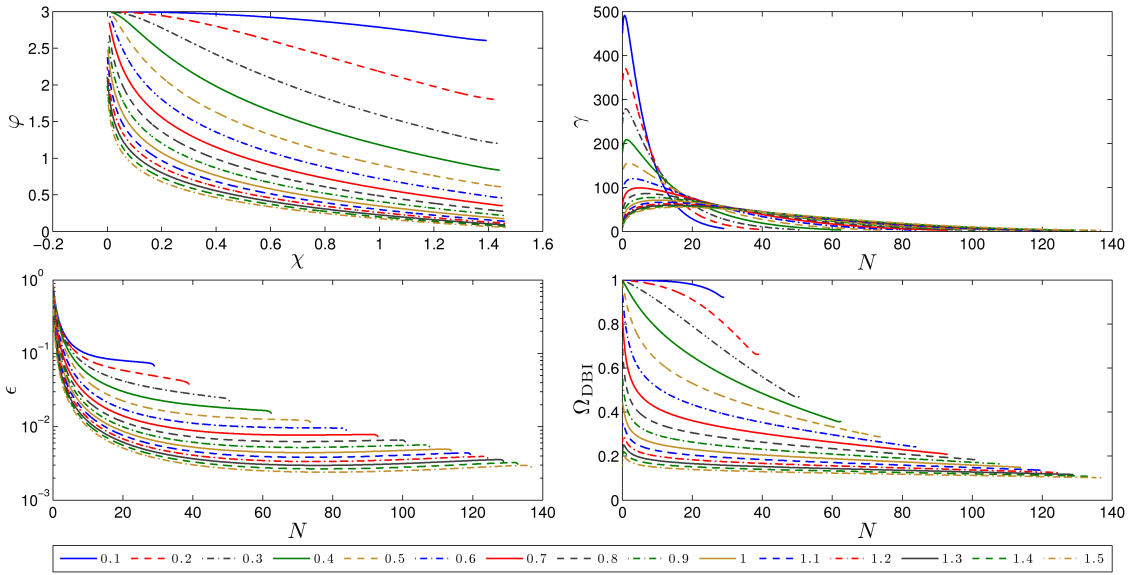}
	\caption{
	The effect of increasing $\beta$ from 0.1 to 1.5 for the offset potential case.
	The upper-left panel shows the trajectory in field space (starting at $\chi=1.5$) with
	$\varphi$ tracking the minimum,
	the upper-right panel shows the DBI boost factor $\gamma$,
	the lower-left panel shows the slow-roll parameter $\eta$ and the
	lower-right panel shows the density parameter of the DBI field. 
	It can be seen that increasing the coupling increases the duration of inflation and decreases the 	
	maximum boost factor.
 	Parameter values: $\lambda = 8\times 10^{12}$, $m=8\times 10^{-6}$, $U_0 = 10^{-10}$, 
	$\mu=0.15$, $\eta=3$.
	}
	\label{fig:NonMinCoupQuad}
\end{figure}

In the non-minimally coupled case, we find numerically that for a wide range of parameters $\varphi$ quickly settles in the minimum of its effective potential, even when substantially displaced (cf. fig. \ref{fig:LargeMAlp}).
This considerably simplifies the analysis as one can make the assumption $\varphi=\varphi_{\rm min}$ for most of the time during inflation,
so the background dynamics are largely independent of the initial conditions for the scalar field $\varphi$.
To better understand the dynamics in this case, consider the condition for the minimum (\ref{eq:MinCond})
for this potential,
\eq \label{eq:QuadMinEqn}
2U_0(\varphi-\eta) \simeq -4\beta A^4 V.
\eqe
Using this, we can rewrite the potentials in terms of $W\equiv W(x)$
with $x\approx 8\beta^2Ve^{4\beta\eta}/U_0$ as
\eq
A^4V \simeq \frac{U_0 W}{8\beta^2} \qquad U = \frac{U_0 W^2}{16\beta^2},
\eqe 
so the Hubble parameter becomes
\eq \label{eq:HQuadApprox}
3H^2 \simeq (1+\tfrac{1}{2}W)A^4 V.
\eqe
One can differentiate (\ref{eq:QuadMinEqn}) with respect to time to get
$\dot\varphi$ in terms of $\dot\chi$
\eq
\dot\varphi \simeq -\frac{1}{2\beta}\left( \frac{W}{1+W} \right)\frac{\dot\chi}{\chi},
\eqe 
where we have used the quadratic form of the $\chi$ potential to write $V_\chi/V  = 2\chi$.
Thus, the rate of change of the scalar field $\dot\varphi$ has an explicit dependence on $\dot\chi$.
Focusing on the rate of change of the Hubble parameter, it can be seen from eqn. (\ref{eq:Fried2}) that for the kinetic term of the DBI field to dominate over that of the scalar field, we need the condition $A^2\gamma\dot\chi^2 \gg \dot\varphi^2$ to be satisfied i.e.
\eqa
A^2\gamma\dot\chi^2 &\gg& \frac{1}{4\beta^2}\left( \frac{W}{1+W} \right)^2\frac{\dot\chi^2}{\chi^2}, \nn\\
\Rightarrow \gamma &\gg& \frac{A^2m^2}{8\beta^2}\left( \frac{W}{1+W} \right)^2 \frac{1}{A^4 V} \nn\\
\Rightarrow \gamma &\gg&\left[ \frac{(2+W)W^2}{48\beta^2(1+W)^2}  \right] \left( \frac{A^2 m^2}{H^2} \right) ,\nn\\
\eqae
where (\ref{eq:HQuadApprox}) was used in the third step. Notice that since $W/4\beta$ is equal to the dislocation of $\varphi$ from its minimum, $W/2$ is fairly small, around $\mathcal{O}(1)$.
Since the effective mass of the DBI field is $mA$ [cf. (\ref{eq:Echi})], which should be (at least) the order of the Hubble scale, we find that the $\varphi$ field is dynamically subdominant (i.e. -$\dot H = \tfrac{1}{2}(\dot\varphi^2+A^2\gamma\dot\chi^2) \approx A^2\gamma\dot\chi^2 $) if $\gamma \gg \mathcal{O}(1)$. 
So, when $\chi$ is affected by the DBI terms in the action, the slow-roll parameter becomes
\eq \label{eq:QuadEpsApprox}
\epsilon \simeq \frac{3A^2\gamma\dot\chi^2}{(2+W)A^4V}\simeq \frac{1}{A^2(1+\tfrac{1}{2}W)}\left( \frac{3\gamma\dot\chi^2}{2V} \right).
\eqe
The case in which $\varphi_{\rm min}>0$ is thus of particular interest as this gives $A\gtrsim1$,
yielding
 \[
 \epsilon \lesssim \left( \frac{3\gamma\dot\chi^2}{2V} \right) \simeq \epsilon_{\rm DBI}
 \]
 where $\epsilon_{\rm DBI}$ is the slow-roll parameter for a standard DBI field\footnote
{
It should be noted that the factor of $A^{-2}$ in the definition of $\gamma$ in the Einstein frame
means that $\gamma$ is smaller in the coupled case compared to the standard DBI scenario so
even discounting the small factor $A^{-2}(1+\tfrac{1}{2}W)^{-1}$ in (\protect\ref{eq:QuadEpsApprox}),
 $\tfrac{3}{2}\gamma\dot\chi^2/V$ is generally less than $\epsilon_{\rm DBI}$.
}
 of mass $m$.
Since $\epsilon$ is inversely related to the number of efolds of inflation $N_{\rm max} = \ln (a_f/a_i)$ by
\[
N_{\rm max}  = \int^{t_f}_{t_i}  H dt = \int^{t_f}_{t_i} \frac{1}{\sqrt{\epsilon}}\sqrt{-\dot H} dt
\simeq \int^{\chi_{\rm ini}}_0 A\sqrt{\frac{\gamma}{2\epsilon}} d\chi,
\]
this results in a increase in the number of efolds of inflation. This is
evident in fig. \ref{fig:NonMinCoupQuad}, in which the
the evolution of the fields is plotted for several values of $\beta$.
One can see that increasing the coupling from 0.1 to 1.5 decreases the slow-roll parameter
by over an order of magnitude, causing the 
the number of efolds 
of inflation to increase from $\sim 30$ to $\sim 140$.
Also, the maximum value of the DBI boost factor is dramatically reduced by a factor of 10, which
could lead to interesting observational deviations from the standard DBI model (discussed in the final section).

If $A\ll 1$, the effective mass of the DBI field is decreased, which has the twofold effect of increasing the slow-roll parameter and decreasing the number of efolds of inflation, as well as affecting the DBI speed limit via the factor of $A^{-2}$ under the square root in eqn. (\ref{eq:gam}).
This can be seen in  the cases presented in fig. \ref{fig:LargeMAlp}. As $\varphi_{\rm min}$ is negative, $A\lesssim1$ and the number of efolds of inflation is decreased compared to the minimally coupled case. 

\begin{figure}[t] 
	\begin{center}
	 	\begin{tabular}{cc}
			   \includegraphics[width=0.48\textwidth]{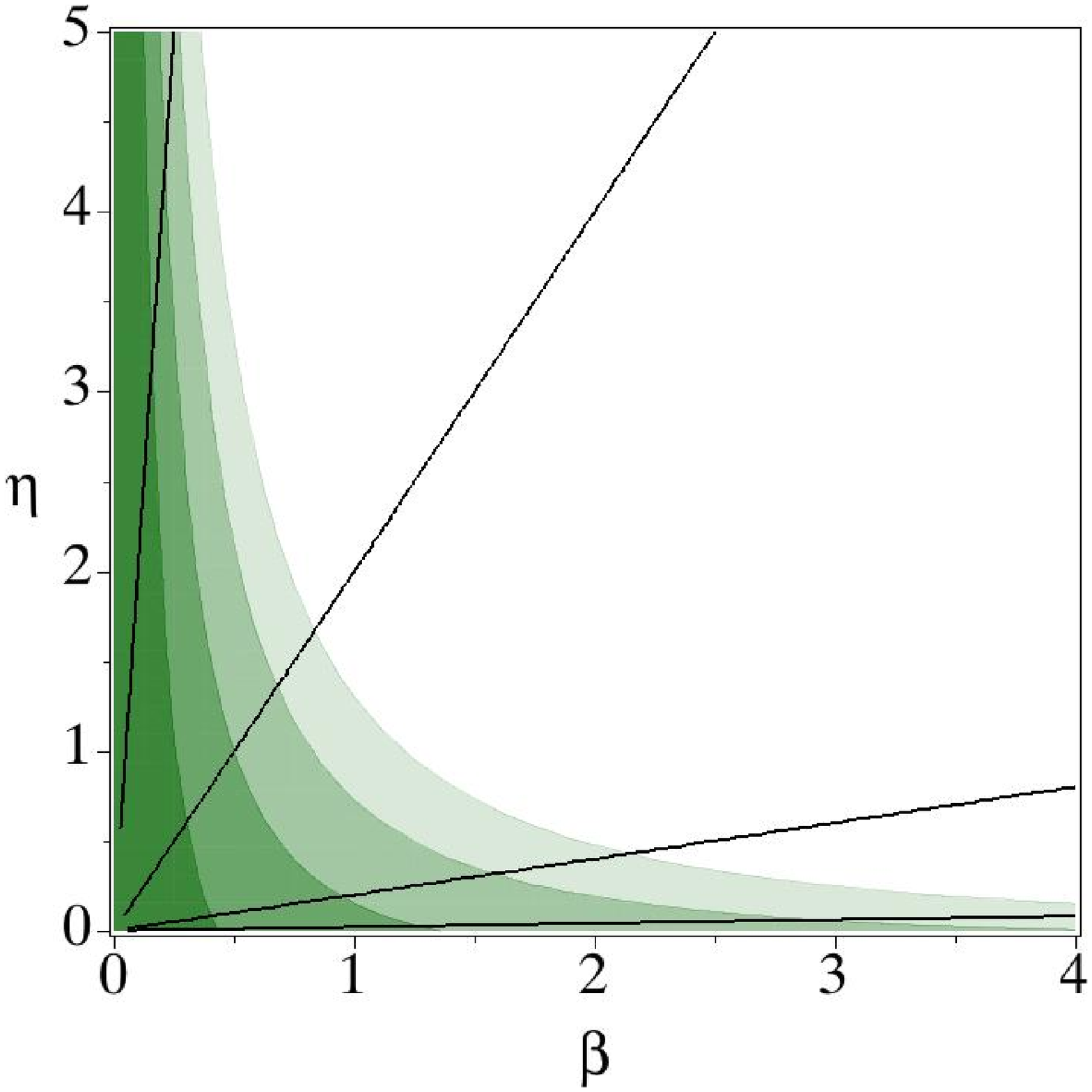} 
			& \includegraphics[width=0.48\textwidth]{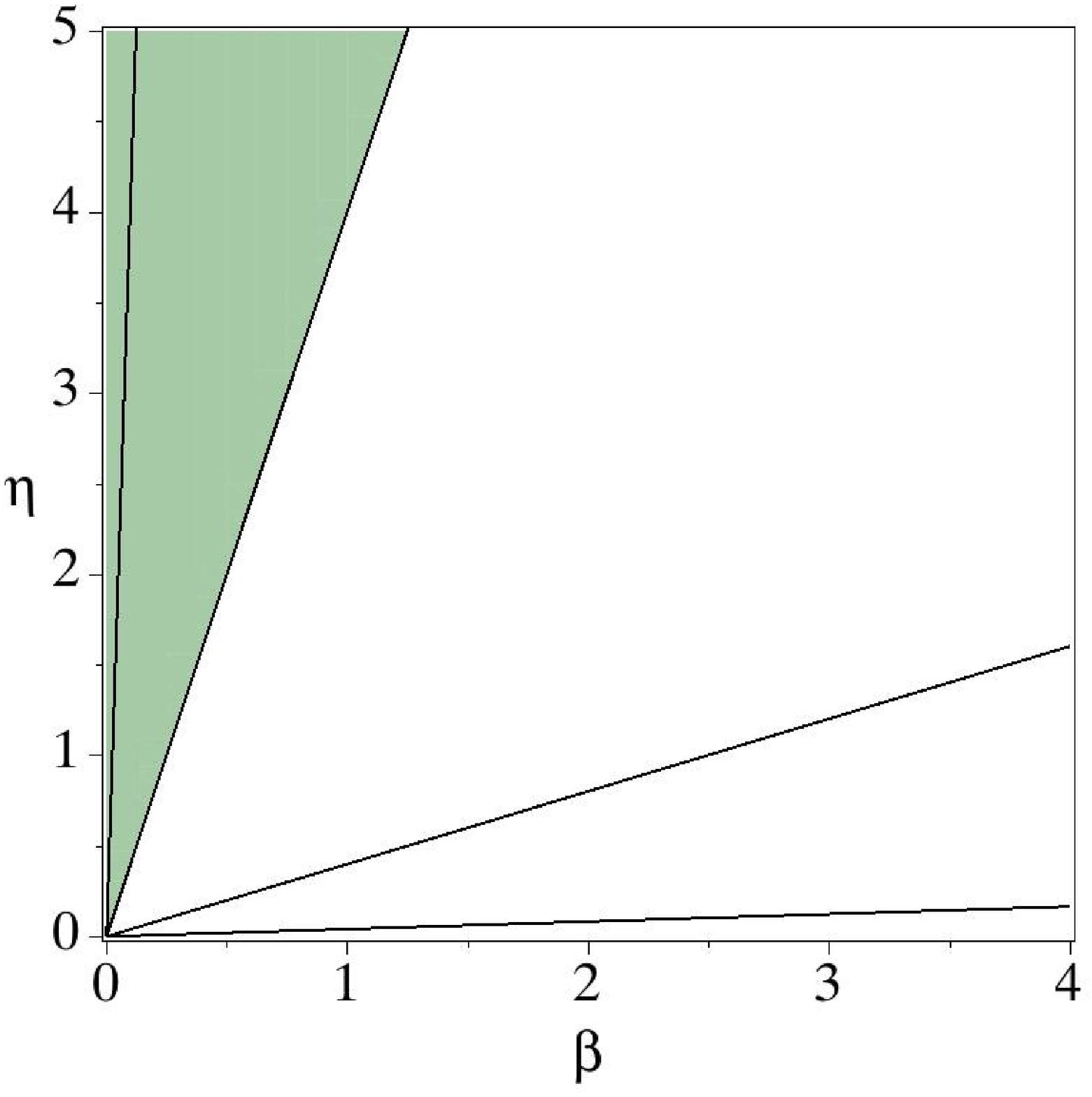} 
		\end{tabular}
	\end{center}
	\caption{
	Conditions for a positive minimum and DBI dominance for the offset quadratic (left panel) and 	
	exponential (right panel) potentials.
	Left panel: regions above the solid lines satisfy the condition for $\varphi_{\rm min}>0$ 
	(\protect\ref{eq:WCond1}) using $V/U_0 = 0.01, 0.1, 1, 10$ (starting from the bottom).  Shaded 
	regions indicate 	
	where the DBI energy is dominant (\protect\ref{eq:WCond2}) with the lightest region corresponding
	 to $V/U_0 = 0.01$.
	 Right panel: the equivalent conditions for the
	 exponential potential. In this case the region satisfying the condition for DBI dominance 
	  (\protect\ref	{eq:LogCond2}),
	  represented by the shaded region,	
	 is independent of the ratio
	 $V/U_0$.
	 }
	 \label{fig:PhiMinLims}
\end{figure}

One can then ask, what conditions need to be fulfilled for in order for the coupling $A$ to be greater than one? 
Substituting (\ref{eq:QuadMin}) into $\varphi_{\rm min}>0$  gives the condition 
\eq \label{eq:WCond1}
W<4\beta\eta,
\eqe
which is plotted in the left-panel of fig. \ref{fig:PhiMinLims} for several values of the ratio $V/U_0$. As one might expect, it is easier to satisfy this condition with larger
values\footnote
{ 
Note that the mexican hat potential with minima at positive and negative values of $\varphi$ is not suitable here as, if the field starts at small values, it is dragged in the direction of the negative minimum by the term on the RHS of eqn. (\ref{eq:varphi}) giving $A<1$.
} 
of the bare minimum $\eta$, as the minimum of the effective potential is shifted to larger (more positive) values.
Using (\ref{eq:HQuadApprox}), the condition for DBI dominance, $\Omega_{\rm DBI}>\Omega_\varphi$ is    
\eq \label{eq:WCond2}
W<2,
\eqe
which is also plotted in  fig. \ref{fig:PhiMinLims}. 
Naively, one might expect the situation wherein the DBI plays a dominant role to be characterised by $V/U_0\gg 1$. However, this means that the second term in (\ref{eq:QuadMin}) is dominant
unless the offset is large
so the net effect is $\varphi_{\rm min}<0$, leading to smaller values of the energy density $\rho_\chi\sim A^4(\varphi)V$. 
One can see in fig. \ref{fig:PhiMinLims} that
the region of the $\beta,\eta$ parameter space that can satisfy both (\ref{eq:WCond1}) and (\ref{eq:WCond2}) is fairly small meaning that for a large number of cases for which $A>1$ the DBI energy density is subdominant. However, this does not mean that $\chi$ is dynamically unimportant: the opposite is the case. The large contribution of the energy density of the scalar field arises as $\varphi$ is trapped in a false vacuum away from its true minimum, so if $\varphi$ tracks the minimum, its evolution is determined solely by the DBI field.
This can be seen in the examples shown in fig. \ref{fig:NonMinCoupQuad}.
In each of the cases presented here $\Omega_{\rm DBI} \lesssim 1$, but the field is never completely 
subdominant. However, despite the significant contribution of $\Omega_\varphi$ to the energy budget, the
dynamics are controlled by the DBI field and $\varphi$ is trapped in the $\chi$-dependent minimum of its effective potential.  (This behaviour is similar to hybrid inflation \cite{Linde:1993cn}, see also \cite{Koh:2010kg} for a recent discussion of hybrid inflation in scalar-tensor theory.)

\subsubsection{Exponential potential}

\begin{figure}[t]
	\centering
	\includegraphics[width=1\textwidth]{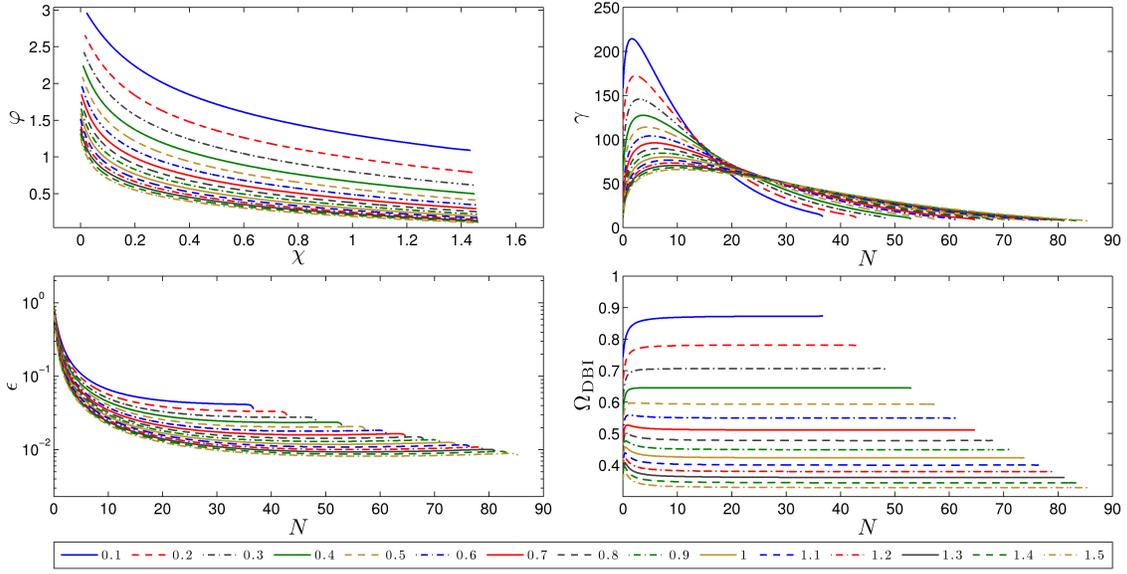}
	\caption{
	The effect of increasing $\beta$ from 0.1 to 1.5 for the exponential potential case.
	Quantities plotted are the same as fig. \protect\ref{fig:NonMinCoupQuad}.
 	Parameter values: $\lambda = 2\times 10^{13}$, $m= 10^{-7}$, $U0 = 5\times5^{-10}$, 
	$\mu=0.32$, $\eta=2.9$.
	}
	\label{fig:NonMinCoupExp}
\end{figure}

As well as the offset quadratic potential, we also consider an exponential potential
\eq
U(\varphi) = U_0 \exp(-\eta\varphi),
\eqe
in which the parameter $\eta$ determines the slope and, under the influence of the bare potential, $\varphi$ moves to larger values.
The condition for the minimum in this case reads
\eq  \label{eq:ExpMinEqn}
U \simeq \frac{4\beta}{\eta}A^4V,
\eqe
so the minimum is given by the logarithmic function
\eq
 \varphi_{\rm min} = \frac{1}{4\beta+\eta}\log\left(\frac{\eta U_0}{4\beta V}\right).
\eqe
Note that in this case, as the fields are dominated by their potentials, the energy densities of the
fields are proportional to each other. Thus, we have a tracker solution for the fields, in which the density parameters are constant $\Omega_{\rm DBI} \propto \Omega_\varphi \simeq {\rm const}$ (as can be seen in fig. \ref{fig:NonMinCoupExp}). 
The condition for $\varphi_{\rm min}>0$ is therefore
\eq \label{eq:LogCond1}
\eta > 4\beta(V/U_0).
\eqe
This is plotted in the right-panel of fig. \ref{fig:PhiMinLims}, and one can see that the
condition is similar to that for the offset quadratic potential, where $\eta$ was the position
of the true minimum. This is simply because in both cases increasing the parameter $\eta$
steepens the bare potential at small values of $\varphi$.

Using (\ref{eq:QuadMinEqn}), the Friedmann equation can be written,
\eq \label{eq:HExpApprox}
3H^2 = \left( 1+4\beta/\eta \right)A^4V,
\eqe
so the condition for DBI dominance is independent of the ratio $V/U_0$:
\eq \label{eq:LogCond2}
\eta > 4\beta.
\eqe
This is also shown in the right-panel of fig. \ref{fig:PhiMinLims}. 
Differentiating (\ref{eq:ExpMinEqn}) with respect to time and repeating the steps in the previous section
yields a simple expression for the rate of change of $\varphi$ in the minimum
\eq
\dot\varphi \simeq -\frac{1}{\eta(1+4\beta\eta)}\frac{\dot\chi}{\chi},
\eqe 
so the condition $\dot\varphi^2 \ll A^2\gamma\dot\chi^2$ implies that
\eq
\gamma \gg \left[ \frac{(1+4\beta/\eta)}{6\eta(1+4\beta\eta)} \right]\left(\frac{A^2m^2}{H^2}  \right)\gtrsim\mathcal{O}(1),
\eqe
must be satisfied, which is true when $\chi$ is affected by the DBI terms in the action.
As with the case the of the offset potential, we find that the slow-roll parameter
\eq
 \label{eq:ExpEpsApprox}
\epsilon \simeq \frac{3A^2\gamma\dot\chi^2}{2(1+4\beta/\eta)A^4V}\simeq 
\frac{1}{A^2(1+4\beta/\eta)}\left( \frac{3\gamma\dot\chi^2}{2V} \right) < \epsilon_{\rm DBI},
\eqe
 is, as in the previous case, smaller than the standard DBI model when $A>1$,  leading to to a greater amount of efolds of 
 inflation (see fig. \ref{fig:NonMinCoupExp}). Also, as  in the example of the offset potential, one can see in fig. \ref{fig:NonMinCoupExp} a decrease in the boost factor of the DBI field.

\section{Cosmological perturbations}
\label{Sec:Perturbations}

\subsection{First order perturbation equations}

We now discuss the linear perturbations in our model. 
The model under consideration is part of a much larger class of multifield models with a Lagrangian 
of the form $P=P(X^{IJ},\phi^{K})$, where $X^{IJ}\equiv-\tfrac{1}{2}\partial_{\mu}\phi^{I}\partial^{\mu}\phi^{J}$ is the kinetic term involving derivatives of the scalar fields $\phi^{I}$ and $\phi^{J}$. 
The equations of motion for the linear perturbations in the general case are calculated in \cite{dbi3}, but 
in order to render transparent the phenomenology that results from the presence of the coupling
and to compare to the standard case, we shall perturb the field equations resulting from the 
action (\ref{eq:Einsteinaction}) in the standard way.
We decompose the fields $\varphi$ and $\chi$ into a homogeneous and perturbed part
\[
\varphi(t,{\bf x}) = \varphi(t) + \delta\varphi(t,{\bf x}),  \qquad  \chi(t,{\bf x}) = \chi(t) + \delta\chi(t,{\bf x}).
\]
We shall work with the Fourier components $\delta\varphi_{\bf k}(t)$ and $\delta\chi_{\bf k}(t)$ so that spatial derivatives can be written $\nabla^2 \delta\varphi = -k^2 \delta\varphi$. Here, as below, we omit the subscript ${\bf k}$ to simplify the notation.
In the longitudinal gauge and in the absence of anisotropic stress, the scalar perturbations of the FRW metric can be expressed as 
\begin{equation} \label{eq:pert_FRW}
ds^2 = -(1+2\Psi)dt^2+a^2(1 - 2\Psi)\delta_{ij}dx^idx^j.
\end{equation} 
The perturbed Einstein equations give 
\begin{eqnarray}
3H(\dot{\Psi}+H\Psi)+\frac{k^2}{a^2}\Psi = -\frac{1}{2M_{\rm Pl}^2}\delta\rho, \\
\dot{\Psi}+H\Psi = -\frac{1}{2M_{\rm Pl}^2}\delta q, \\
\ddot{\Psi}+4H\dot{\Psi}+(3H^2+2\dot{H})\Psi = \frac{1}{2M_{\rm Pl}^2}\delta p,
\end{eqnarray}
where $\delta\rho$, $\delta q$ and $\delta p$ are perturbations of the total energy density, momentum potential and pressure respectively, given by
\begin{eqnarray}
\delta q &=& -\left[ \dot{\varphi}(\delta \varphi)+{\gamma}A^2\dot{\chi}(\delta\chi) \right], 
\label{eq:delta_q} \\
\delta\rho &=& [\dot{\varphi}(\delta \varphi)\dot{\ }-\dot{\varphi}^2\Psi] 
+A^2\gamma^3[\dot{\chi}(\delta \chi)\dot{\ }-\dot{\chi}^2\Psi] 
  +\left[ {U}_{\varphi}+4\beta \rho_{\rm DBI}-\beta A^2{\gamma}^3 \dot\chi^2  \right](\delta\varphi)+  \nonumber\\
&{ }&+\left[ A^4 V_\chi+\tfrac{1}{2}A^4\frac{f_\chi}{f^2}(2-3{\gamma}+{\gamma}^{3}) \right](\delta\chi),
\label{eq:delta_rho} \\
\delta p  &=& [\dot{\varphi}(\delta \varphi)\dot{\ }-\dot{\varphi}^2\Psi] 
+A^2\gamma[\dot{\chi}(\delta \chi)\dot{\ }-\dot{\chi}^2\Psi]+
\left[ -{U}_{\varphi}+4\beta  p_{\rm DBI}-\beta A^2{\gamma} \dot\chi^2  \right](\delta\varphi) -  \nonumber\\
&{ }&  -\left[ A^4 V_\chi+\tfrac{1}{2}A^4\frac{f_\chi}{f^2}(2-{\gamma}^{-1}-{\gamma}) \right](\delta\chi).
\label{eq:delta_p}
\end{eqnarray}
The perturbed field equations are
\begin{eqnarray}
(\delta\varphi)\ddot{ \ } &+& 3 H (\delta\varphi)\dot{ \ }
+ \left[ \frac{k^2}{a^2} + U_{\varphi\varphi}- 4\beta^2 {T}_{DBI}
+\beta^2A^2(3c_s^{-1}-c_s^{-3})\dot\chi^2
 \right](\delta\varphi) - \nn \\
 &-& \beta A^2(3c_s^{-1}-c_s^{-3})\dot\chi(\delta\chi)\dot{\ } 
+\beta A^4\left[ \frac{f_\chi}{f^2}\left(4-\tfrac{3}{2}c_s-3 c_s^{-1}+\tfrac{1}{2}c_s^{-3} \right) +4V_\chi \right](\delta\chi)+ \nn \\
&+&  \left[ 2{U}_{\varphi}-2\beta{T}_{DBI}  +\beta A^2(3c_s^{-1}-c_s^{-3})\dot\chi^2\right]\Psi
 -4\dot{\varphi}\dot\Psi  \nn\\
 &{ }& =0
\end{eqnarray}
and
\begin{eqnarray}
(\delta\chi)\ddot{\ } &+&
 \left[ 3 H+2\beta\dot{\varphi}-3\frac{\dot{c_s}}{c_s} \right](\delta\chi)\dot{\ }
+ \Bigg[ c_s^2\frac{k^2}{a^2}+A^2c_s^3V_{\chi\chi}
-\frac{f_\chi}{f}\frac{\dot c_s}{c_s}\dot\chi
-\tfrac{1}{2}\dot\chi^2 c_s f_\chi V_\chi + \nn\\
&{ }& + \tfrac{1}{2}A^2 (1-c_s)^2 \left( c_s \left(\frac{f_\chi}{f^2}\right)_{,\chi} +(1+c_s)f^{-1}\left(\frac{f_\chi}{f}\right)_{,\chi} \right)
 \Bigg](\delta\chi) +
\nn\\
&+& \beta(3c_s^2-1)\dot\chi(\delta\varphi)\dot{\ }
+\beta\left[ A^2 V_{\chi}c_s(1+c_s^2)+A^2\frac{f_\chi}{f^2}c_s(1-c_s)^2+2\frac{\dot c_s}{c_s}\dot\chi \right](\delta\varphi) - \nn\\ 
&-&(3 c_s^{2}+1)\dot\chi\dot\Psi
+\left[ A^2 V_{\chi}c_s(1+c_s^2)+A^2\frac{f_\chi}{f^2}c_s(1-c_s)^2+2\frac{\dot c_s}{c_s}\dot\chi \right]\Psi 
 \nn \\
&{ }& =  0
\end{eqnarray}
where we have rewritten factors of $\gamma$ using the expression for the sound speed for the DBI field $c_s^2=\gamma^{-2}$. In terms of the gauge invariant variables,
\eqa
Q_\varphi \equiv (\delta\varphi)+\frac{\dot{\varphi}}{ H}\Psi, \\
Q_\chi \equiv (\delta\chi)+\frac{\dot\chi}{ H}\Psi,
\eqae
the equations form a closed system,
\eqa
\ddot{Q}\varphi &+& 3 H\dot{Q}\varphi+B_\varphi\dot Q_\chi+\left(\frac{k^2}{ a^2}+C_{\varphi\varphi}\right) Q_\varphi+C_{\varphi\chi} Q_\chi =0, \label{eq:dot_Q_varphi}\\
\ddot{Q}\chi &+& \left[ 3 H+2\beta\dot{\varphi}-3\frac{\dot c_s}{c_s} \right]\dot{Q}\chi +B_\chi\dot Q_\varphi+\left(\frac{k^2}{ a^2}c^2_s+C_{\chi\chi}\right) Q_\chi+C_{\chi\varphi} Q_\varphi =0
\label{eq:dot_Q_chi}
\eqae
where we have defined the following coefficients
\eqa
B_\chi &=&  \left[\beta  (3 c_s^{2}-1) - \frac{\dot{\varphi}}{ 2H}(1-c_s^{2})
 \right] \dot\chi, \\ 
B_\varphi &=&   -A^{2}c_s^{-3} B_\varphi, \\
\nn\\%
C_{\varphi\varphi}&=&
\beta \left( \frac{\dot{{\varphi}}}{ H} \right)A^4 f^{-1} (3c_s+1)(1-c_s^{-1})^3
-\beta^2 A^4 f^{-1}(16-8c_s^{-1}-9c_s+c_s^{-3})+
\nonumber 
\\ %
&{}& 
+3\dot{\varphi}^2 - c_s^{-3}(1+c_s^2)A^2\frac{\dot{\varphi}^2\dot\chi^2}{4 H^2} 
-\frac{\dot{\varphi}^4}{2  H^2}
+ \frac{2\dot{\varphi} V_{T,\varphi}}{ H}
+V_{T,\varphi\varphi},
\\ %
C_{\varphi\chi}&=& 
\frac{A^4\dot{\varphi}}{4 H}\frac{f_\chi}{f^2}c_s(1-c_s^{-1})^2(c_s^{-2}+2c_s^{-1}-1)
+3c_s^{-1}A^2\dot\chi\dot{\varphi}
-c_s^{-4}(1+c_s^2)A^4\frac{\dot{\varphi}\dot\chi^3}{4 H^2}-
\nn\\ %
&{}& 
-\tfrac{1}{2}\beta A^4 f^{-1}(3c_s+1)(1-c_s^{-1})^3\left[ \frac{f_\chi}{f} -\frac{A^2\dot\chi}{ H c_s}\right]
-c_s^{-1}A^2\frac{\dot{\varphi}^3\dot\chi}{2 H^2}
+ \frac{V_{T,\chi}\dot{\varphi}}{ H} + \nn \\
&{}&
+\frac{c_s^{-1}A^2\dot{\chi}}{ H}V_{T,\varphi}
+V_{T,\varphi\chi},
\\ %
C_{\chi\chi}&=& 
\frac{A^4\dot\chi}{ H}\frac{f_\chi}{f^2}(1-c_s)^2
-\left[ \frac{f_\chi}{f} -\frac{A^2\dot\chi}{ H c_s}\right]\frac{\dot{c_s}}{c_s}\dot\chi
-\tfrac{1}{2}c_s f_\chi A^{-4} \dot\chi^2 V_{T,\chi}  + 
\nn\\ %
&{}& 
+\tfrac{1}{2}A^2 (1-c_s)^2 \left[ c_s \left(\frac{f_\chi}{f^2}\right)_{,\chi} +(1+c_s)f^{-1}\left(\frac{f_\chi}{f}\right)_{,\chi} \right]
+\tfrac{3}{2}A^2\dot\chi^2 c_s^{-1}(1+c_s^2)-
\nn\\ %
&{}& 
-A^4 c_s^{-2}\frac{\dot\chi^4}{2 H^2}
-A^2 c_s^{-1}(1+c_s^2)\frac{\dot\chi^2\dot{\varphi}^2}{4 H^2}
+\frac{\dot\chi V_{T,\chi}}{ H}(1+c_s^2)
+c_s^{3}A^{-2} V_{T,\chi\chi},
\\ %
C_{\chi\varphi}&=& 
\left(2\beta+\frac{\dot{\varphi}}{ H}\right)
\left[ \tfrac{1}{2}A^2\frac{f_\chi}{f^2}c_s(1-c_s)^2 +\frac{\dot{c_s}}{c_s}\dot\chi  \right]
-2\beta\frac{A^4\dot\chi}{ H}f^{-1}(1-c_s)^2  
-c_s^{-1}\frac{A^2\dot{\varphi}\dot\chi^3}{2 H^2}+
\nn\\ %
&{}& 
+ \tfrac{1}{2}(1+c_s^2)\left[
3\dot{\varphi}\dot\chi
-\frac{\dot{\varphi}^3\dot\chi}{2 H^2}
+2 \beta c_sA^{-2} V_{T,\chi}
+c_s \frac{\dot{\varphi}A^{-2} V_{T,\chi}}{ H}
+\frac{V_{T,\varphi}\dot\chi}{ H}
 \right].
\eqae

In these expressions, $V_T(\varphi,\chi) = U(\varphi) + A^4(\varphi)V(\chi)$ is the total potential. Comparing the form of these equations to \cite{Lalak:2007vi}, where the authors treated the case of two coupled canonical fields, there are marked similarities. The principle differences are due to factors of $c_s$
that arise due to the non-canonical kinetic term of the DBI field. 
Let us define the auxiliary fields $\nu_\varphi$ and $\nu_\chi$, by
\eqa
\nu_\varphi = r_\varphi Q_\varphi, \qquad \nu_\chi = r_\chi Q_\chi,
\eqae
with $r_\chi = aAc_s^{-3/2}$ and $r_\varphi=a$,
which correspond to the perturbation variables used in the analysis of the perturbations in single-field canonical and  DBI inflation respectively. The equations of motion (\ref{eq:dot_Q_varphi}) and (\ref{eq:dot_Q_varphi}) can then be written
\eqa
\nu_\varphi'' - B\nu_\chi' + \left[ k^2 + a^2C_{\varphi\varphi} - \frac{r_\varphi''}{r_\varphi} \right]\nu_\varphi
+ \left[ \left( \frac{r_\varphi}{r_\chi} \right)a^2C_{\varphi\chi} + B\frac{r_\chi'}{r_\chi}  \right]\nu_\chi &=& 0,
\\ %
\nu_\chi'' + B\nu_\varphi' + \left[ k^2c_s^2 + a^2C_{\chi\chi} - \frac{r_\chi''}{r_\chi} \right]\nu_\chi
+ \left[  \left( \frac{r_\chi}{r_\varphi} \right)a^2C_{\chi\varphi} - B\frac{r_\varphi'}{r_\varphi}  \right]\nu_\varphi &=& 0,
\eqae
with $B = r_\chi B_\chi$. In this form the symmetry between the equations is more clear: each field perturbation exchanges energy with the other via the friction term $B$ as well as the mass terms.
The obvious complexity of these equations makes an analytical treatment difficult, so in the following section we shall integrate the perturbation equations numerically. 

\subsection{The evolution of the perturbations}

\begin{figure}[t] 
	\centering
	 \includegraphics[width=1\textwidth]{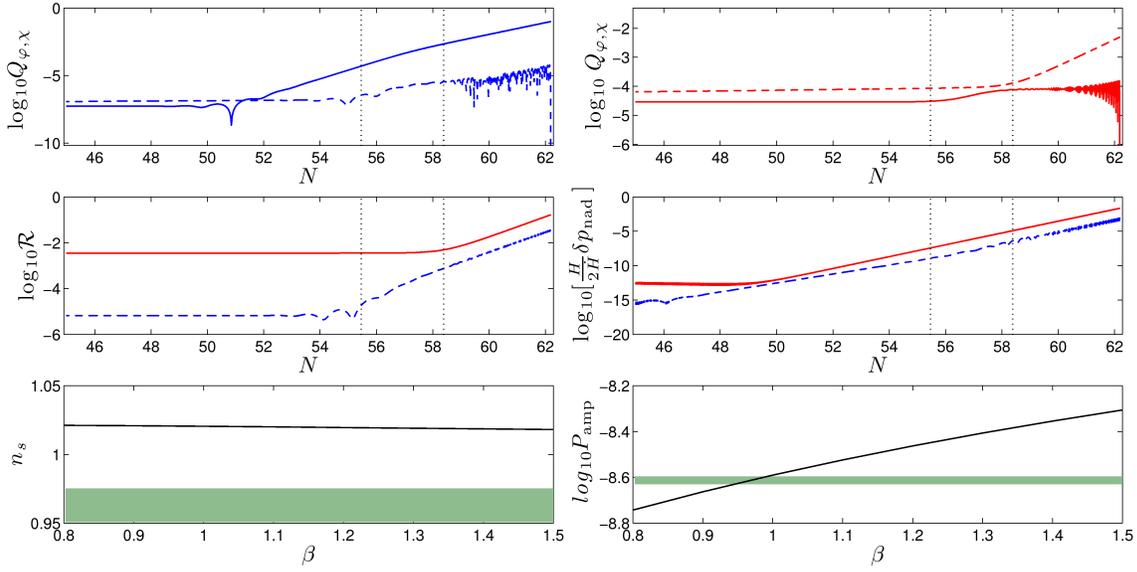}	
	 \caption{
	Perturbation quantities for the exponential case with $\gamma\gg 1$ shown in fig. 
	\protect\ref{fig:NonMinCoupExp}. The upper panels show the evolution of the amplitude 
	of the variables $Q_{\varphi}$ (solid) and $Q_{\chi}$ (dashed) for the first run (top-left) 
	with $Q_{\chi}^{\rm ini}=0$ and
	the second run (top-right) with $Q_{\varphi}^{\rm ini}=0$ for the pivot scale with $\beta=1$.
	 (Perturbations are normalised so that
	 $|Q_{\varphi}^{\rm ini}|=0.1$ for the first run.) The centre plots  show the evolution 
	of the comoving curvature perturbation $\mathcal{R}$ (centre-left) and the non-adiabatic contribution
	to its time derivative $H/(2\dot H)\delta p_{\rm nad}$ for the first (blue, dashed) and second 
	(red, solid) runs in this case. Vertical dotted lines indicate the sound and curvature horizons for
	 this mode.
	The lower panels show the variation in the resulting value of the spectral index $n_s$ (bottom-left)
	and power spectrum amplitude $P_{\rm amp}$ (bottom-right) with the coupling $\beta$. 
	The shaded regions shows observational values of these quantities (WMAP+ BAO+H0) 
	at $68\%$ c.l \protect\cite{Komatsu:2010fb}.
	 }
	 \label{fig:LargeGamPert}
\end{figure}

In order to integrate the perturbation equations, it is important to specify the initial conditions.
In the case of single-field k-inflation, the approach introduced by Garriga and Mukhanov \cite{Garriga:1999vw} is to construct the auxiliary variable $\nu = z \mathcal{R}$ , with $z = a(\rho+p)^{1/2}/(c_s H)$. One then quantises  $\nu$, substituting $\nu\propto \nu_k e^{i\bf k \cdot x}$ to obtain,
\eq
 \frac{d^2 v_k}{d \tau^2} + \left( c_s^2 k^2 -\frac{1}{z}\frac{d^2z}{d\tau^2} \right)v_k =0.
\eqe
 Assuming the quantum fluctuations are in the trivial vacuum state, the solutions for $k^2c_s^2 \gg a^2H^2$ are oscillatory as in the standard case, but with a frequency reduced by a factor of $c_s$ that also affects the normalisation:
\eq
\nu_k = \frac{1}{\sqrt{2kc_s}}e^{-i k c_s\tau}.
\label{eq:con_BD}
\eqe

The above procedure can be applied in the single field case, but when there are two fields present, one has to be to careful when choosing initial conditions for the numerical run. As in the standard two-field case we can neglect correlations between the perturbations for
modes well within the horizon, so the short wavelength solutions for the auxiliary variables $\nu_\varphi$ and $\nu_\chi$ are
\eq \label{eq:nu_ini}
\nu_\varphi = \frac{1}{\sqrt{2k}}e^{-i k \tau}, \qquad
\nu_\chi =  \frac{1}{\sqrt{2kc_s}}e^{-i k c_s\tau},
\eqe 
so the relative normalisation of the two perturbations is dependent on the sound speed of the DBI field.
However, in order to avoid introducing correlations between the fields at early times
when integrating numerically (in which case the Bunch-Davies vacua would be inappropriate)
the perturbations must be integrated twice, and the results combined to form the 
curvature power spectrum (see Appendix for a more in-depth description of the numerical method).
In the first run, shown in the upper-left panel of fig. \ref{fig:LargeGamPert}, the scalar field perturbations
are set in the Bunch-Davies vacuum with $Q_\chi^{\rm ini}=0$ and in the second run (upper-right panel of fig. \ref{fig:LargeGamPert}) $Q_\varphi^{\rm ini}=0$ and the perturbations of the DBI field are set in the trivial vacuum.

The comoving curvature perturbation is given by,
\eq
\mathcal{R} = \Psi - \frac{H}{\rho+p}\delta q
\eqe
where $\delta q$ is the momentum potential given in eq. (\ref{eq:delta_q}) and $p$ and $\rho$ are the total background pressure and energy density respectively. In terms of the gauge invariant variables $Q_\varphi$ and $Q_\chi$ this yields the simple expression,
\eq \label{eq:ComovingR}
\mathcal{R} = \frac{H}{(-2\dot H)}\left[ \dot\varphi Q_\varphi + A^2\gamma\dot\chi Q_\chi  \right].
\eqe
where $(-2\dot H) = \dot\varphi^2+A^2\gamma\dot\chi^2 $.

When the fields are minimally coupled and both fields are evolving, as in the standard DBI case, the sound speed $c_s$ affects the evolution of the perturbations so that $Q_\chi$ freezes in at the sound horizon ($kc_s = aH$) while $Q_\varphi$ freezes in at the curvature horizon $k=aH$. 
However, in the non-minimally coupled case, discussed in Sec. \ref{sec:NonMinCoup}, the homogeneous field $\varphi$ is in the minimum of its effective potential and its perturbations are suppressed, so the
primary contribution to the comoving curvature perturbation $\mathcal{R}$ (middle-left in fig. \ref{fig:LargeGamPert}) is the second run, in which the $Q_\chi$ is dominant.
Thus in the coupled case, the curvature perturbation freezes in at the sound horizon of the DBI field, as in a 
single-field DBI system.
The lower panels in fig.  \ref{fig:LargeGamPert} show the variation in the resulting value of the spectral 
 index $n_s$ (bottom-left) and power spectrum amplitude $P_{\rm amp}$ (bottom-right) with the 
 coupling $\beta$. In single field DBI, the power spectrum amplitude increases with 
 the number of efolds of inflation. Here, the increase in $\beta$ is accompanied by an increase in $N_{\rm max}$, leading to a monotonic increase of  $P_{\rm amp}$ with $\beta$.
 The spectral index is much affected by the small sound speed of the DBI field in this example,
  which counteracts the 
 red-tilt due to the deviation from a pure de Sitter stage, so that $n_s$ exhibits a slightly blue tilt.

\begin{figure}[t] 
	\centering
	 \includegraphics[width=1\textwidth]{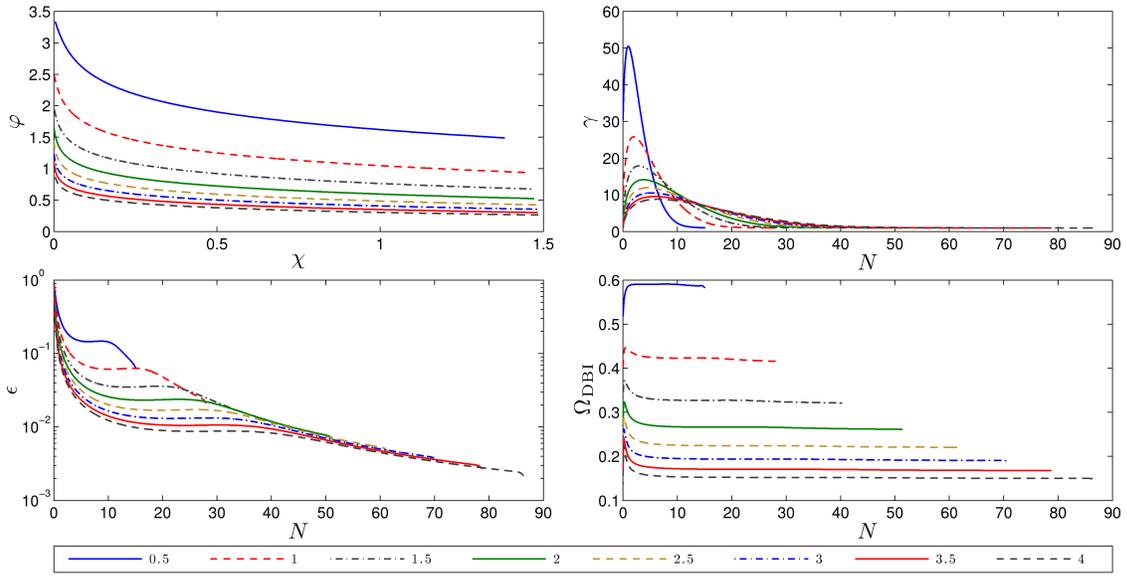}	\caption{
	 Background dynamics with small $\gamma$ (exponential potential). In this case, 
	 as $\beta$ increases from 0.5 to 4, the large coupling 
	 decreases the warp factor so $\chi$ slow-rolls before reaching smaller values so $\gamma\approx 1$
	 for a considerable period. 
	Quantities plotted are the same as fig. \protect\ref{fig:NonMinCoupExp}.
 	Parameter values: $\lambda = 7.8\times 10^{11}$, $m= 2.8\times10^{-6}$, $U_0 = 7\times10^{-9}$, 
	$\mu=0.07$, $\eta=2.8$.
	 }
	 \label{fig:SmallGamBG}
\end{figure}

\begin{figure}[t] 
	\centering
	 \includegraphics[width=1\textwidth]{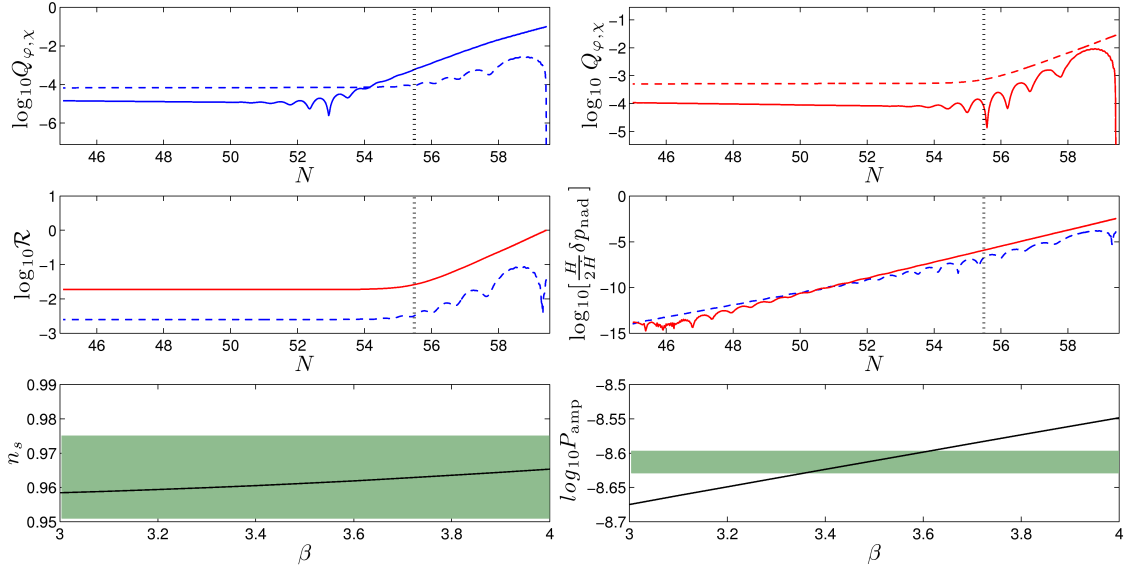}	\caption{
	 The evolution of the perturbation quantities for the small $\gamma$ case 
	 plotted in fig. \protect\ref{fig:SmallGamBG}
	 for the pivot scale with $\beta=3.5$. 
	 Quantities plotted are the same as in fig. \protect\ref{fig:LargeGamPert}. Note that,
	 as $\gamma\approx 1$, the sound and curvature horizons (dotted, vertical lines) are
	 indistinguishable and modes freeze-on as in slow-roll inflation, giving a red-tilted spectrum.
	}
	\label{fig:SmallGamPert}
\end{figure}

If the coupling is large, or the mass of the DBI field small, the value of $\gamma$ can be close to 1 
when the modes that affect the CMB freeze in. An example of this is shown in fig. \ref{fig:SmallGamBG},
where increasing the value of $\beta$ not only decreases $\gamma$ and increases $N_{\rm max}$ but gives rise to a period of slow-roll at the start of the run. This can be seen by comparing the 
behaviour of the slow-roll parameter $\epsilon$ and the boost factor $\gamma$ at the start and end of the run. At first $\gamma\sim 1$; afterwards increasing to values $\lesssim 20$ when the effective DBI warp factor $A^2f$ becomes significant. 
In this case, the sound and curvature horizons are almost identical and the field perturbations behave
as in the case of canonical fields, freezing in at $k=aH$ (see fig. \ref{fig:SmallGamPert}). As in the previous case,
the perturbations of the $\varphi$ field are subdominant, and only give small corrections to $\mathcal{R}$.
It can be seen in fig. \ref{fig:SmallGamPert}, that, due to this effect, the spectral index displays a red-tilt
putting it within current observational limits.

 The rate of change of the curvature perturbation can be written,
\begin{equation}
\dot{\mathcal{R}} = \frac{H}{(-2\dot H)}\left[ (\delta p_{nad}) + \frac{\dot p}{\dot\rho} (\delta \rho_m)  \right]  
\label{eq:dot_R}
\end{equation}
where non-adiabatic pressure is defined by,
\begin{equation}
(\delta p_{nad})  \equiv  (\delta p) -\frac{\dot p}{\dot\rho} (\delta \rho).
\end{equation}
The quantity $\delta \rho_m \equiv \delta\rho-3H\delta q$ is the gauge invariant comoving density perturbation. This appears on the RHS of the Poisson equation,
\eq
\frac{k^2}{a^2}\Psi = -\tfrac{1}{2}\delta \rho_m,
\eqe
and thus decays in the long-wavelength/late-time limit. It is given explicitly by,
\eqa\hspace{-2em}
\delta \rho_m &=& \left[\dot\varphi\dot Q_\varphi +A^2\gamma^3\dot\chi\dot Q_\chi \right] 
+ \frac{2\dot H}{H}\left[ \frac{1}{2H}(\dot\varphi^2+A^2\gamma^3\dot\chi^2)-3H    \right]\mathcal{R} \nn\\
&{}& +\left[ V_{T, \varphi} - \beta A^4f^{-1}(4-5\gamma+\gamma^3) \right]Q_\varphi 
+\left[ V_{T, \chi} + \tfrac{1}{2}\beta A^4 \frac{f_\chi}{f^2}(2-3\gamma+\gamma^3) \right]Q_\chi.
\label{eq:delta_rho_m}
\eqae

In the standard two-field case, the non-adiabatic pressure perturbation consists of a contribution proportional to $\delta\rho_m$ and another proportional to the entropy field perturbation $\delta s$. The former is negligible at late times, but the latter can act as a source for the curvature perturbation even after the fluctuations grow larger than the curvature horizon. In order to obtain an explicit form for $\delta p _{\rm nad}$, we can rewrite (\ref{eq:delta_rho_m}) as
\eqa
\delta \rho_m &=& -\frac{2\dot H}{H}\dot{ \mathcal{R}} + A^2 (\gamma^3-\gamma)\dot\chi\dot Q_\chi
+\left[ A^2\dot\chi^2\frac{\dot H}{H^2} (\gamma^3-\gamma) - \frac{2\ddot H}{H} \right]\mathcal{R} 
\nn\\
&{}& + \left[ \beta A^2 f^{-1} (4\gamma-3\gamma^{-1}-\gamma^3 - 2\ddot\varphi  \right]Q_\varphi
+ \left[ \tfrac{1}{2}f_\chi\gamma^3\dot\chi^4 - 2(A^2\gamma\dot\chi)\dot{ }  \right]Q_\chi
\eqae
and combine with (\ref{eq:dot_R}) to get,
\eq \label{eq:deltapnad}
\delta p_{\rm nad} = \left[1-\frac{\dot p}{\dot \rho}\right]\delta\rho_m
+ \frac{2}{\sqrt{-2\dot H}}\left[ \sqrt{-2\dot H} P_s +\beta A^3 \gamma^{3/2} \dot\chi^3 \right](\delta s)
+\left[\frac{\dot\gamma}{\sqrt{-2\dot H}}Q_\sigma - Q_\gamma  \right]A^2\dot\chi^2.
\eqe
Here, we have introduced adiabatic and entropic fields, defined as
\eqa
Q_\sigma &=& \frac{1}{\sqrt{-2\dot H}}\left[ \dot\varphi Q_\varphi + A^2\gamma\dot\chi Q_\chi \right] 
\label{eq:adiabatic}\\
\delta s &=& \frac{A\gamma^{1/2}}{\sqrt{-2\dot H}} \left[ \dot\varphi Q_\chi - \dot\chi Q_\varphi  \right]
\label{eq:entropic}
\eqae 
in a similar manner to the coupled two-field system \cite{DiMarco:2002eb},  so that the adiabatic field
is related to the comoving curvature perturbation by
\eq
\mathcal{R} = \frac{H}{\sqrt{-2\dot H}}Q_\sigma,
\eqe
and the partial derivative of the total pressure with respect to the entropy field is,
\eqa
P_s  &=&  \frac{1}{\sqrt{-2\dot H}}\left[ \frac{\dot\varphi}{A\gamma^{1/2}}\frac{\p p}{\p\chi} - A\gamma^{1/2}\dot\chi \frac{\p p}{\p\varphi}\right] \nn\\
&=&    \frac{1}{\sqrt{-2\dot H}} [  V_{T , \varphi}A\gamma^{1/2}\dot\chi  - \frac{\dot\varphi}{A\gamma^{1/2}}V_{T,\chi}
+\tfrac{1}{2}A^3\frac{f_\chi}{f^2}\gamma^{1/2}(1-\gamma^{-1})^2\dot\varphi \nn\\
&{}&
-\beta A^5 f^{-1}\gamma^{1/2}(4-3\gamma^{-1}-\gamma))\dot\chi ]. 
\eqae
An interesting feature of eqn. (\ref{eq:deltapnad}) is that, unlike the canonical two-field case and the standard coupled case, this expression cannot be decomposed into purely entropic and adiabatic perturbation terms. The extra terms are expressed in the gauge-invariant combination $Q_\gamma$,
\eqa
Q_\gamma &=& \delta\gamma + \frac{\dot\gamma}{H}\Psi \nn\\
&=& \gamma^3A^{-2}f\dot\chi\dot Q_\chi + (\gamma^3-\gamma)\left( \tfrac{1}{2} \frac{f_\chi}{f} Q_\chi 
-\beta Q_\varphi - \frac{\sqrt{-2\dot H}}{2H} Q_\sigma \right).
\label{eq:delta_gamma}
\eqae
The fact that this term appears like this suggests that the field decomposition is not appropriate
in this case. We can also see this if we consider the form of the equation of motion for $Q_\sigma$:
\eqa
\ddot Q_\sigma &=& \frac{1}{\sqrt{-2\dot H}}\left[ \dot\varphi \ddot Q_\varphi + A^2\gamma\dot\chi \ddot Q_\chi \right] + ... \nn\\
&=& - \frac{k^2}{a^2}\frac{1}{\sqrt{-2\dot H}}\left[ \dot\varphi Q_\varphi + A^2\gamma^{-1}\dot\chi  Q_\chi \right] + ... 
\eqae
where the $...$ represent terms in $\dot Q\varphi$ and $\dot Q\chi$ as well as terms proportional to $Q_\varphi$ and $Q_\chi$ not proportional to $k^2/a^2$. From this, we can see that the $\ddot Q_\sigma$ equation cannot be recast in the form, $\ddot x + c_s^2\tfrac{k^2}{a^2}x + ... = 0$, as a scalar perturbation can. The same is true of the entropy field.
Even if the definition of the adiabatic field were changed to include a non-canonical kinetic term itself, i.e.
\eq
\dot\sigma\sqrt{P_Z} Q_\sigma = \dot\varphi \sqrt{P_X}Q_\varphi + \dot\chi\sqrt{P_Y} Q_\chi,
\eqe
where $P_Z\dot\sigma^2\equiv-2\dot H$, no extra terms proportion to $k^2/a^2$ would arise, so the original conclusion would stand. The description of non-adiabatic pressure perturbations in such a system
as this could perhaps be better treated by the formalism introduced in \cite{Malik:2004tf}, considering separate potential and kinetic fluids. However, as in this paper we focus on the coupled case in which
the perturbations of $\varphi$ are suppressed, non-adiabatic perturbations are negligible.
We will thus discuss this issue in more detail in future work focusing on the non-minimally coupled case discussed briefly in Sec. \ref{sec:NonMinCoup}.

In the non-minimally coupled case under consideration, we have found that the coupled DBI model
is effectively a single-field system, and so the contribution of non-adiabatic fields is negligible. 
This can be seen in the middle-right panels of figs. \ref{fig:LargeGamPert} and \ref{fig:SmallGamPert}, in
which the non-adiabatic pressure contribution to $\dot{\mathcal{R}}$, $\frac{H}{2\dot H}(\delta p_{\rm nad )}$, is shown to decrease exponentially as inflation proceeds. Thus, the curvature perturbation is constant
to a high degree of accuracy on large scales.

\section{Discussion} 
\label{Sec:Conclusions}

In this paper we have studied the consequences of coupling the DBI action to 
a canonical scalar field in a scalar-tensor theory. The motivation for this work is 
purely phenomenological: the DBI inflationary scenario has its motivation in 
string theory, but it is an interesting example of k-inflation in its own right. The 
additional scalar field might have its origin as a moduli field in a higher-dimensional 
setup, but we leave the explicit construction for future work (although, see \cite{Brax:2009hd} on how DBI inflation could be related to scalar tensor theories). 

We have investigated the case in which the canonical scalar field is heavy 
during inflation. We found this to be the case even for couplings of order a tenth of the 
strength of gravity ($\beta\approx 0.1$). Thus, we believe this to be a realistic scenario. With couplings larger than this, the additional scalar field is forced into the minimum of its effective potential. Thus, the parameters of the DBI field, which are dependent on the additional scalar field, vary during inflation. As a consequence, we find that the number of e-folds is extended and the boost factor is decreased (when compared to standard DBI inflation with the same bare parameters). The energy density during inflation comes mostly from the additional scalar field that sits in the minimum of the effective potential, which varies slowly as the DBI field evolves. But because the canonical field is heavy during inflation, its perturbations are negligible and only the DBI perturbations contribute to the curvature perturbation. Since the boost factor is smaller than in the standard case, the level of non-Gaussianity (which is proportional to $\gamma^2$) is smaller than the standard DBI case. In fact, we found that the presence of the canonical scalar field alleviates the problems of the DBI inflationary scenario. 

We have considered two potentials for the canonical scalar field: a quadratic potential (with a minimum $\varphi_{\rm min}>0$) and an exponential potential. 
The difference between the phenomenology of these two models is slight, mainly due to the presence of a minimum in the bare potential in the offset quadratic case, which affects the background dynamics towards the end of inflation, when the DBI contribution to the effective potential is less important. 
In the case of the exponential potential, the energy densities of the scalar fields (dominated by their potential energies) are proportional to each other (as in a tracker solution). 
In both cases we find that the behaviour of the perturbations is extremely similar, affected only indirectly by presence of the $\varphi$ field in its minimum and the details of the DBI dynamics; thus, we venture to suggest that our conclusions may hold for any choice of potential steep enough to allow $\varphi$ to closely track the minimum of its effective potential.

Finally, we have investigated two possible couplings. 
In the minimally coupled case we find find that if one fine-tunes the mass scales so that both fields contribute roughly equally to the total energy density for a considerable number of efolds, 
the dynamics of the DBI field can be significantly affected by the additional contribution to the Hubble damping, leading to smaller values of $\gamma$ (relative to the DBI dominant case). 
As both fields are evolving, the interplay between the field perturbations suggests that
isocurvature fluctuations could be important.
However, the minimally coupled case is extremely sensitive to small changes in the field masses  and the initial conditions of both fields.
In the non-minimally coupled case, we find that the coupled DBI model
is effectively a single-field system, and so the contribution of non-adiabatic fields is negligible. 
Moreover, the curvature perturbation is constant to a high degree of accuracy on large scales.

We have not discussed in detail the range of allowed parameter for the model but will study this question in future work.  A range of parameters affect the prediction for the spectral index (for both the exponential and the quadratic potential) and we are currently scanning the parameter space numerically. 
Much has to be learned in this model, such as the decoupling limit $\beta \rightarrow 0$. In this case, the canonical scalar field will become light and isocurvature modes become important. We expect that the signatures of non-gaussianity will be very different.

\begin{acknowledgments}
We are grateful to Philippe Brax for useful discussions and comments on an earlier draft of this paper. CvdB is partly supported by STFC. DFM thanks the Research Council of Norway for the FRINAT grant 197251/V30. DFM is also partially supported by project CERN/FP/109381/2009 and PTDC/FIS/102742/2008. JW is supported by EPSRC and a Yggdrasil Grant from the research council of Norway, and would like to thank the members of the Institute of Theoretical Astrophysics at the University of Norway for their kind hospitality while parts of this work were completed. 
\end{acknowledgments}

\appendix
\section{Numerical method}
\label{appendix}

In this short appendix, we devote a few words to an explanation of the general numerical method
used to integrate the perturbations. 
The pivot scale for the CMB ($k_{\rm pivot} = 0.002 \text{Mpc}^{-1}$ \cite{Komatsu:2010fb}) at which the power spectrum amplitude is calculated is identified with the mode that crosses the curvature horizon ($k=aH$) 55 efolds before the end of inflation (cf. \cite{Liddle:2003as}), when the slow-roll parameter $\epsilon$ ceases to be smaller than 1. The spectral index is calculated by integrating the perturbation equations for $k$-values in the range $k\in[k_{\rm pivot}\times 10^{-2}, k_{\rm pivot}\times 10^{2}]$, starting when $k c_s=50aH$ (so the wavelength of the perturbation is much smaller than the sound horizon of the DBI field) and ending 45 efolds from the end of inflation, at which time the modes are beyond the horizon and have frozen in.
We do not integrate the equations (\ref{eq:dot_Q_varphi}) and (\ref{eq:dot_Q_chi}) with respect to conformal time or $N=\log a$, as is commonly done,
but instead use logarithmic time $x=\log t$, which is efficient in dealing with both the background and perturbation variables.
In this case the oscillatory initial conditions (\ref{eq:nu_ini}) for $Q_\varphi$ and $Q_\chi$ are
\eq \label{eq:Q_ICs}
Q_\varphi = \frac{1}{L^{3/2}}\frac{1}{a\sqrt{2k}}e^{-i\frac{k}{a} t},
\qquad Q_\chi = \frac{1}{L^{3/2}}\frac{c_s}{aA}\frac{1}{\sqrt{2k}}e^{-i\frac{k}{a}c_s t},
\eqe
where $L$ is a normalisation factor chosen so that $Q_\varphi^{\rm ini}=0.1$. Representing $d/dx$ by $'$ and writing $\mathcal{H} = a'/a$ the derivatives are
\eq
Q_{\varphi}' = -\left(\mathcal{H}+i\frac{k}{a}t\right)Q_\varphi, \qquad
Q_{\chi}' = -\left(\mathcal{H}+\beta\varphi' - \frac{c_s'}{c_s}+i\frac{k}{a}c_s t\right)Q_\chi,
\eqe
where as in the previous expressions, background quantities are evaluated at $t_{\rm ini}=\exp(x_{\rm ini})$.

As in the standard two-field case (cf .\cite{Tsujikawa:2002qx}) in order to make sure that the modes are uncorrelated
deep within the horizon, we perform two numerical runs. the In the first run $Q_\varphi$
is set in the Bunch-Davies vacuum (\ref{eq:Q_ICs}) with $Q_\chi^{\rm ini}=0$ and in the second run $Q_\varphi^{\rm ini}=0$ and the perturbations of the DBI field are set in the (quasi) Bunch-Davies vacuum. Evaluating the curvature perturbation (\ref{eq:ComovingR}) for each run yields the solutions 
$\mathcal{R}_1$ and $\mathcal{R}_2$, which can be combined to form the curvature power spectrum
\eq
\mathcal{P}_{\mathcal{R}} = \frac{k^3}{2\pi^2}\left( |\mathcal{R}_1|^2 + |\mathcal{R}_2|^2 \right).
\eqe
The power spectrum amplitude (for the pivot mode $k_{\rm pivot}$) and the spectral index can then be calculated at the end of the numerical run.

\bibliographystyle{JHEP}
\bibliography{CoupledDBI_refs}

\newcommand{\noopsort}[1]{} \newcommand{\singleletter}[1]{#1}
\providecommand{\href}[2]{#2}\begingroup\raggedright\begin{thebibliography}{10}

\bibitem{McAllister:2007bg}
L.~McAllister and E.~Silverstein, {\it {String Cosmology: A Review}},  {\em
  Gen. Rel. Grav.} {\bf 40} (2008) 565--605,
  [\href{http://xxx.lanl.gov/abs/0710.2951}{{\tt arXiv:0710.2951}}].

\bibitem{Baumann:2009ni}
D.~Baumann and L.~McAllister, {\it {Advances in Inflation in String Theory}},
  {\em Ann. Rev. Nucl. Part. Sci.} {\bf 59} (2009) 67--94,
  [\href{http://xxx.lanl.gov/abs/0901.0265}{{\tt arXiv:0901.0265}}].

\bibitem{Copeland:1994vg}
E.~J. Copeland, A.~R. Liddle, D.~H. Lyth, E.~D. Stewart, and D.~Wands, {\it
  {False vacuum inflation with Einstein gravity}},  {\em Phys.Rev.} {\bf D49}
  (1994) 6410--6433, [\href{http://xxx.lanl.gov/abs/astro-ph/9401011}{{\tt
  astro-ph/9401011}}].

\bibitem{Stewart:1994ts}
E.~D. Stewart, {\it {Inflation, supergravity and superstrings}},  {\em
  Phys.Rev.} {\bf D51} (1995) 6847--6853,
  [\href{http://xxx.lanl.gov/abs/hep-ph/9405389}{{\tt hep-ph/9405389}}].

\bibitem{eta1}
J.~McDonald, {\it {F-term hybrid inflation, the eta-problem and extra
  dimensions}},  {\em JHEP} {\bf 12} (2002) 029,
  [\href{http://xxx.lanl.gov/abs/hep-ph/0201016}{{\tt hep-ph/0201016}}].

\bibitem{Easson:2009kk}
D.~A. Easson and R.~Gregory, {\it {Circumventing the eta problem in building an
  inflationary model in string theory}},  {\em Phys. Rev.} {\bf D80} (2009)
  083518, [\href{http://xxx.lanl.gov/abs/0902.1798}{{\tt arXiv:0902.1798}}].

\bibitem{eta3}
S.~Antusch, M.~Bastero-Gil, K.~Dutta, S.~F. King, and P.~M. Kostka, {\it
  {Solving the $\eta$-Problem in Hybrid Inflation with Heisenberg Symmetry and
  Stabilized Modulus}},  {\em JCAP} {\bf 0901} (2009) 040,
  [\href{http://xxx.lanl.gov/abs/0808.2425}{{\tt arXiv:0808.2425}}].

\bibitem{Silverstein:2003hf}
E.~Silverstein and D.~Tong, {\it {Scalar Speed Limits and Cosmology:
  Acceleration from D- cceleration}},  {\em Phys. Rev.} {\bf D70} (2004)
  103505, [\href{http://xxx.lanl.gov/abs/hep-th/0310221}{{\tt
  hep-th/0310221}}].

\bibitem{Alishahiha:2004eh}
M.~Alishahiha, E.~Silverstein, and D.~Tong, {\it {DBI in the sky}},  {\em Phys.
  Rev.} {\bf D70} (2004) 123505,
  [\href{http://xxx.lanl.gov/abs/hep-th/0404084}{{\tt hep-th/0404084}}].

\bibitem{ArmendarizPicon:1999rj}
C.~Armendariz-Picon, T.~Damour, and V.~F. Mukhanov, {\it {k-Inflation}},  {\em
  Phys. Lett.} {\bf B458} (1999) 209--218,
  [\href{http://xxx.lanl.gov/abs/hep-th/9904075}{{\tt hep-th/9904075}}].

\bibitem{Garriga:1999vw}
J.~Garriga and V.~F. Mukhanov, {\it {Perturbations in k-inflation}},  {\em
  Phys. Lett.} {\bf B458} (1999) 219--225,
  [\href{http://xxx.lanl.gov/abs/hep-th/9904176}{{\tt hep-th/9904176}}].

\bibitem{k2}
F.~Helmer and S.~Winitzki, {\it {Self-reproduction in k-inflation}},  {\em
  Phys.Rev.} {\bf D74} (2006) 063528,
  [\href{http://xxx.lanl.gov/abs/gr-qc/0608019}{{\tt gr-qc/0608019}}].

\bibitem{k3}
G.~Panotopoulos, {\it {Detectable primordial non-gaussianities and
  gravitational waves in k-inflation}},  {\em Phys.Rev.} {\bf D76} (2007)
  127302, [\href{http://xxx.lanl.gov/abs/0712.1713}{{\tt arXiv:0712.1713}}].

\bibitem{k4}
V.~Taveras and N.~Yunes, {\it {The Barbero-Immirzi Parameter as a Scalar Field:
  K-Inflation from Loop Quantum Gravity?}},  {\em Phys.Rev.} {\bf D78} (2008)
  064070, [\href{http://xxx.lanl.gov/abs/0807.2652}{{\tt arXiv:0807.2652}}].

\bibitem{k5}
N.~Bose and A.~Majumdar, {\it {A k-essence Model Of Inflation, Dark Matter and
  Dark Energy}},  {\em Phys.Rev.} {\bf D79} (2009) 103517,
  [\href{http://xxx.lanl.gov/abs/0812.4131}{{\tt arXiv:0812.4131}}].

\bibitem{k6}
Y.-b. Yue and B.~Chen, {\it {K-Inflation with a Dark Energy Coupling}},  {\em
  Phys.Rev.} {\bf D81} (2010) 023506,
  [\href{http://xxx.lanl.gov/abs/0906.0145}{{\tt arXiv:0906.0145}}].

\bibitem{k7}
N.~Bose and A.~Majumdar, {\it {Unified Model of k-Inflation, Dark Matter and
  Dark Energy}},  {\em Phys.Rev.} {\bf D80} (2009) 103508,
  [\href{http://xxx.lanl.gov/abs/0907.2330}{{\tt arXiv:0907.2330}}].

\bibitem{k8}
C.~Ringeval, {\it {Dirac-Born-Infeld and k-inflation: the CMB anisotropies from
  string theory}},  {\em J.Phys.Conf.Ser.} {\bf 203} (2010) 012056,
  [\href{http://xxx.lanl.gov/abs/0910.2167}{{\tt arXiv:0910.2167}}]. * Brief
  entry *.

\bibitem{Bartolo:2004if}
N.~Bartolo, E.~Komatsu, S.~Matarrese, and A.~Riotto, {\it {Non-Gaussianity from
  inflation: Theory and observations}},  {\em Phys. Rept.} {\bf 402} (2004)
  103--266, [\href{http://xxx.lanl.gov/abs/astro-ph/0406398}{{\tt
  astro-ph/0406398}}].

\bibitem{Liguori:2010hx}
M.~Liguori, E.~Sefusatti, J.~Fergusson, and E.~Shellard, {\it {Primordial
  non-Gaussianity and Bispectrum Measurements in the Cosmic Microwave
  Background and Large-Scale Structure}},  {\em Adv.Astron.} {\bf 2010} (2010)
  980523, [\href{http://xxx.lanl.gov/abs/1001.4707}{{\tt arXiv:1001.4707}}]. *
  Temporary entry *.

\bibitem{DiMarco:2002eb}
F.~Di~Marco, F.~Finelli, and R.~Brandenberger, {\it {Adiabatic and Isocurvature
  Perturbations for Multifield Generalized Einstein Models}},  {\em Phys. Rev.}
  {\bf D67} (2003) 063512,
  [\href{http://xxx.lanl.gov/abs/astro-ph/0211276}{{\tt astro-ph/0211276}}].

\bibitem{Lalak:2007vi}
Z.~Lalak, D.~Langlois, S.~Pokorski, and K.~Turzynski, {\it {Curvature and
  isocurvature perturbations in two-field inflation}},  {\em JCAP} {\bf 0707}
  (2007) 014, [\href{http://xxx.lanl.gov/abs/0704.0212}{{\tt
  arXiv:0704.0212}}].

\bibitem{dbi10}
S.~Renaux-Petel, {\it {Primordial bispectrum and trispectrum in multifield
  Dirac- Born-Infeld inflation}},
  \href{http://xxx.lanl.gov/abs/1004.3693}{{\tt arXiv:1004.3693}}.

\bibitem{dbi9}
S.~Mizuno, F.~Arroja, and K.~Koyama, {\it {On the full trispectrum in
  multi-field DBI inflation}},  {\em Phys. Rev.} {\bf D80} (2009) 083517,
  [\href{http://xxx.lanl.gov/abs/0907.2439}{{\tt arXiv:0907.2439}}].

\bibitem{dbi1}
D.~A. Easson, R.~Gregory, D.~F. Mota, G.~Tasinato, and I.~Zavala, {\it
  {Spinflation}},  {\em JCAP} {\bf 0802} (2008) 010,
  [\href{http://xxx.lanl.gov/abs/0709.2666}{{\tt arXiv:0709.2666}}].

\bibitem{dbi3}
D.~Langlois, S.~Renaux-Petel, D.~A. Steer, and T.~Tanaka, {\it {Primordial
  perturbations and non-Gaussianities in DBI and general multi-field
  inflation}},  {\em Phys. Rev.} {\bf D78} (2008) 063523,
  [\href{http://xxx.lanl.gov/abs/0806.0336}{{\tt arXiv:0806.0336}}].

\bibitem{dbi2}
C.~P. Burgess, R.~Easther, A.~Mazumdar, D.~F. Mota, and T.~Multamaki, {\it
  {Multiple inflation, cosmic string networks and the string landscape}},  {\em
  JHEP} {\bf 05} (2005) 067,
  [\href{http://xxx.lanl.gov/abs/hep-th/0501125}{{\tt hep-th/0501125}}].

\bibitem{dbi7}
S.~Renaux-Petel, {\it {Combined local and equilateral non-Gaussianities from
  multifield DBI inflation}},  {\em JCAP} {\bf 0910} (2009) 012,
  [\href{http://xxx.lanl.gov/abs/0907.2476}{{\tt arXiv:0907.2476}}].

\bibitem{dbi4}
D.~Langlois, S.~Renaux-Petel, D.~A. Steer, and T.~Tanaka, {\it {Primordial
  fluctuations and non-Gaussianities in multi- field DBI inflation}},  {\em
  Phys. Rev. Lett.} {\bf 101} (2008) 061301,
  [\href{http://xxx.lanl.gov/abs/0804.3139}{{\tt arXiv:0804.3139}}].

\bibitem{dbi8}
D.~Langlois, S.~Renaux-Petel, and D.~A. Steer, {\it {Multi-field DBI inflation:
  introducing bulk forms and revisiting the gravitational wave constraints}},
  {\em JCAP} {\bf 0904} (2009) 021,
  [\href{http://xxx.lanl.gov/abs/0902.2941}{{\tt arXiv:0902.2941}}].

\bibitem{Cai:2009hw}
Y.-F. Cai and H.-Y. Xia, {\it {Inflation with multiple sound speeds: a model of
  multiple DBI type actions and non-Gaussianities}},  {\em Phys. Lett.} {\bf
  B677} (2009) 226--234, [\href{http://xxx.lanl.gov/abs/0904.0062}{{\tt
  arXiv:0904.0062}}].

\bibitem{Kachru:2003sx}
S.~Kachru {\em et.~al.}, {\it {Towards inflation in string theory}},  {\em
  JCAP} {\bf 0310} (2003) 013,
  [\href{http://xxx.lanl.gov/abs/hep-th/0308055}{{\tt hep-th/0308055}}].

\bibitem{Easson:2009wc}
D.~A. Easson, S.~Mukohyama, and B.~A. Powell, {\it {Observational Signatures of
  Gravitational Couplings in DBI Inflation}},  {\em Phys. Rev.} {\bf D81}
  (2010) 023512, [\href{http://xxx.lanl.gov/abs/0910.1353}{{\tt
  arXiv:0910.1353}}].

\bibitem{Brax:2009hd}
P.~Brax and E.~Cluzel, {\it {Brane Bremsstrahlung in DBI Inflation}},  {\em
  JCAP} {\bf 1003} (2010) 016, [\href{http://xxx.lanl.gov/abs/0912.0806}{{\tt
  arXiv:0912.0806}}].

\bibitem{Brax:2010tq}
P.~Brax and E.~Cluzel, {\it {Trapped Brane Features in DBI Inflation}},
  \href{http://xxx.lanl.gov/abs/1010.4462}{{\tt arXiv:1010.4462}}. * Temporary
  entry *.

\bibitem{Cai:2010wt}
Y.-F. Cai, J.~B. Dent, and D.~A. Easson, {\it {Warm DBI Inflation}},
  \href{http://xxx.lanl.gov/abs/1011.4074}{{\tt arXiv:1011.4074}}. * Temporary
  entry *.

\bibitem{Kecskemeti:2006cg}
S.~Kecskemeti, J.~Maiden, G.~Shiu, and B.~Underwood, {\it {DBI inflation in the
  tip region of a warped throat}},  {\em JHEP} {\bf 09} (2006) 076,
  [\href{http://xxx.lanl.gov/abs/hep-th/0605189}{{\tt hep-th/0605189}}].

\bibitem{Kinney:2002qn}
W.~H. Kinney, {\it {Inflation: Flow, fixed points and observables to arbitrary
  order in slow roll}},  {\em Phys. Rev.} {\bf D66} (2002) 083508,
  [\href{http://xxx.lanl.gov/abs/astro-ph/0206032}{{\tt astro-ph/0206032}}].

\bibitem{Tzirakis:2008qy}
K.~Tzirakis and W.~H. Kinney, {\it {Non-canonical generalizations of slow-roll
  inflation models}},  {\em JCAP} {\bf 0901} (2009) 028,
  [\href{http://xxx.lanl.gov/abs/0810.0270}{{\tt arXiv:0810.0270}}].

\bibitem{Spalinski:2007dv}
M.~Spalinski, {\it {On Power Law Inflation in DBI Models}},  {\em JCAP} {\bf
  0705} (2007) 017, [\href{http://xxx.lanl.gov/abs/hep-th/0702196}{{\tt
  hep-th/0702196}}].

\bibitem{Chimento:2007es}
L.~P. Chimento and R.~Lazkoz, {\it {Bridging geometries and potentials in DBI
  cosmologies}},  {\em Gen. Rel. Grav.} {\bf 40} (2008) 2543--2555,
  [\href{http://xxx.lanl.gov/abs/0711.0712}{{\tt arXiv:0711.0712}}].

\bibitem{Bessada:2009pe}
D.~Bessada, W.~H. Kinney, and K.~Tzirakis, {\it {Inflationary potentials in DBI
  models}},  {\em JCAP} {\bf 0909} (2009) 031,
  [\href{http://xxx.lanl.gov/abs/0907.1311}{{\tt arXiv:0907.1311}}].

\bibitem{Barnaby:2004gg}
N.~Barnaby, C.~P. Burgess, and J.~M. Cline, {\it {Warped reheating in
  brane-antibrane inflation}},  {\em JCAP} {\bf 0504} (2005) 007,
  [\href{http://xxx.lanl.gov/abs/hep-th/0412040}{{\tt hep-th/0412040}}].

\bibitem{Lachapelle:2008sy}
J.~Lachapelle and R.~H. Brandenberger, {\it {Preheating with Non-Standard
  Kinetic Term}},  {\em arXiv:0808.0936} (2008)
  [\href{http://xxx.lanl.gov/abs/0808.0936}{{\tt arXiv:0808.0936}}].

\bibitem{Bouatta:2010bp}
N.~Bouatta, A.-C. Davis, R.~H. Ribeiro, and D.~Seery, {\it {Preheating in
  Dirac-Born-Infeld inflation}},  {\em JCAP} {\bf 1009} (2010) 011,
  [\href{http://xxx.lanl.gov/abs/1005.2425}{{\tt 1005.2425}}].

\bibitem{Chen:2006nt}
X.~Chen, M.-x. Huang, S.~Kachru, and G.~Shiu, {\it {Observational signatures
  and non-Gaussianities of general single field inflation}},  {\em JCAP} {\bf
  0701} (2007) 002, [\href{http://xxx.lanl.gov/abs/hep-th/0605045}{{\tt
  hep-th/0605045}}].

\bibitem{Bean:2007hc}
R.~Bean, S.~E. Shandera, S.~H. Henry~Tye, and J.~Xu, {\it {Comparing Brane
  Inflation to WMAP}},  {\em JCAP} {\bf 0705} (2007) 004,
  [\href{http://xxx.lanl.gov/abs/hep-th/0702107}{{\tt hep-th/0702107}}].

\bibitem{Peiris:2007gz}
H.~V. Peiris, D.~Baumann, B.~Friedman, and A.~Cooray, {\it {Phenomenology of
  D-Brane Inflation with General Speed of Sound}},  {\em Phys. Rev.} {\bf D76}
  (2007) 103517, [\href{http://xxx.lanl.gov/abs/0706.1240}{{\tt
  arXiv:0706.1240}}].

\bibitem{Baumann:2006cd}
D.~Baumann and L.~McAllister, {\it {A Microscopic Limit on Gravitational Waves
  from D-brane Inflation}},  {\em Phys. Rev.} {\bf D75} (2007) 123508,
  [\href{http://xxx.lanl.gov/abs/hep-th/0610285}{{\tt hep-th/0610285}}].

\bibitem{Gordon:2000hv}
C.~Gordon, D.~Wands, B.~A. Bassett, and R.~Maartens, {\it {Adiabatic and
  entropy perturbations from inflation}},  {\em Phys. Rev.} {\bf D63} (2000)
  023506, [\href{http://xxx.lanl.gov/abs/astro-ph/0009131}{{\tt
  astro-ph/0009131}}].

\bibitem{GrootNibbelink:2001qt}
S.~Groot~Nibbelink and B.~van Tent, {\it {Scalar perturbations during multiple
  field slow-roll inflation}},  {\em Class.Quant.Grav.} {\bf 19} (2002)
  613--640, [\href{http://xxx.lanl.gov/abs/hep-ph/0107272}{{\tt
  hep-ph/0107272}}].

\bibitem{Linde:1993cn}
A.~D. Linde, {\it {Hybrid inflation}},  {\em Phys. Rev.} {\bf D49} (1994)
  748--754, [\href{http://xxx.lanl.gov/abs/astro-ph/9307002}{{\tt
  astro-ph/9307002}}].

\bibitem{Koh:2010kg}
S.~Koh and M.~Minamitsuji, {\it {Non-minimally coupled hybrid inflation}},
  \href{http://xxx.lanl.gov/abs/1011.4655}{{\tt arXiv:1011.4655}}.

\bibitem{Komatsu:2010fb}
E.~Komatsu {\em et.~al.}, {\it {Seven-Year Wilkinson Microwave Anisotropy Probe
  (WMAP) Observations: Cosmological Interpretation}},
  \href{http://xxx.lanl.gov/abs/1001.4538}{{\tt arXiv:1001.4538}}.

\bibitem{Malik:2004tf}
K.~A. Malik and D.~Wands, {\it {Adiabatic and entropy perturbations with
  interacting fluids and fields}},  {\em JCAP} {\bf 0502} (2005) 007,
  [\href{http://xxx.lanl.gov/abs/astro-ph/0411703}{{\tt astro-ph/0411703}}].

\bibitem{Liddle:2003as}
A.~R. Liddle and S.~M. Leach, {\it {How long before the end of inflation were
  observable perturbations produced?}},  {\em Phys.Rev.} {\bf D68} (2003)
  103503, [\href{http://xxx.lanl.gov/abs/astro-ph/0305263}{{\tt
  astro-ph/0305263}}].

\bibitem{Tsujikawa:2002qx}
S.~Tsujikawa, D.~Parkinson, and B.~A. Bassett, {\it {Correlation - consistency
  cartography of the double inflation landscape}},  {\em Phys.Rev.} {\bf D67}
  (2003) 083516, [\href{http://xxx.lanl.gov/abs/astro-ph/0210322}{{\tt
  astro-ph/0210322}}].

\end{thebibliography}\endgroup

\end{document}